%% file: dSph_Jfactors.tex
\documentclass[10pt,aps,prd,preprintnumbers,showpacs,nofootinbib,superscriptaddress,notitlepage,twocolumn,longbibliography]{revtex4-1}

\usepackage{graphicx}
\usepackage{color}
\usepackage{amsmath}
\usepackage{amssymb}
\usepackage{enumerate}
\usepackage{pgf}
\usepackage{multirow}
\usepackage[colorlinks=true,citecolor=blue,urlcolor=blue]{hyperref}
\usepackage{notes2bib}

\newcommand{\PRE}[1]{{#1}} 

\begin{document}

\title{Effective \textit{J}-factors for Milky Way dwarf spheroidal galaxies with velocity-dependent annihilation}

\author{Kimberly K.~Boddy}
\affiliation{\mbox{Department of Physics \& Astronomy,
Johns Hopkins University, Baltimore, Maryland 21218, USA}}

\author{Jason Kumar}
\affiliation{\mbox{Department of Physics \& Astronomy,
University of Hawaiʻi, Honolulu, Hawaiʻi 96822, USA}} 

\author{Andrew B.~Pace}
\affiliation{{Department of Physics and Astronomy,
Mitchell Institute for Fundamental Physics and Astronomy,
Texas A\&M University, College Station, Texas  77843, USA}}
\affiliation{McWilliams Center for Cosmology, Carnegie Mellon University, 5000 Forbes Avenue, Pittsburgh, Pennsylvania 15213, USA}

\author{Jack Runburg}
\affiliation{\mbox{Department of Physics \& Astronomy,
University of Hawaiʻi, Honolulu, Hawaiʻi 96822, USA}}

\author{Louis E.~Strigari\PRE{\vspace*{.1in}}}
\affiliation{{Department of Physics and Astronomy,
Mitchell Institute for Fundamental Physics and Astronomy,
Texas A\&M University, College Station, TX  77843, USA}}

\begin{abstract}
\PRE{\vspace*{.1in}}
We calculate the effective $J$-factors, which determine the strength of indirect detection signals from dark matter annihilation, for 25 dwarf spheroidal galaxies (dSphs). We consider several well-motivated assumptions for the relative velocity dependence of the dark matter annihilation cross section: $\sigma_A v$: $s$-wave (velocity independent), $p$-wave ($\sigma_A v \propto v^2$), $d$-wave ($\sigma_A v \propto v^4$), and Sommerfeld-enhancement in the Coulomb limit ($\sigma_A v \propto 1/v$). As a result we provide the largest and most updated sample of $J$-factors for velocity-dependent annihilation models.  For each scenario, we use Fermi-LAT gamma-ray data to constrain the annihilation cross section. Due to the assumptions made in our gamma-ray data analysis, our bounds are comparable to previous bounds on both the $p$-wave and Sommerfeld-enhanced cross sections using dSphs. Our bounds on the $d$-wave cross section are the first such bounds using indirect detection data.
\end{abstract}

\maketitle

\section{Introduction}

One of the most promising strategies for the indirect detection of dark matter (DM) is the search for gamma rays arising from DM annihilation in dwarf spheroidal galaxies (dSphs).
These targets are especially useful because they have large dark-to-luminous mass ratios, large expected DM annihilation rates, and no standard astrophysical sources of gamma rays.

The flux of gamma rays arising from DM annihilation depends on the properties of the astrophysical source through the $J$-factor.
Under the standard assumption of velocity-independent DM annihilation, the $J$-factor is determined by the DM density profile of the dSph.
If, however, the annihilation cross section is velocity dependent, the calculation of the $J$-factor must account for this velocity dependence by incorporating the full DM velocity distribution~\cite{Robertson:2009bh,Ferrer:2013cla,Boddy:2017vpe,Zhao2017,Petac:2018gue,Boddy:2018ike,Lacroix:2018qqh,Boddy:2019wfg}.
Previous works have estimated these effective $J$-factors for some dSphs, using a variety of techniques, under the assumptions of Sommerfeld-enhanced DM annihilation in the Coulomb limit ($\sigma_A v \propto 1/v$)~\cite{Zhao2016,Boddy:2017vpe,Petac:2018gue,Lacroix:2018qqh,Bergstrom2018} and $p$-wave annihilation ($\sigma_A v \propto v^2$)~\cite{Zhao2016,Zhao2017}.

In this work, we calculate the effective $J$-factors for 25 dSphs of the Milky Way (MW), under well-motivated annihilation models: $s$-wave, $p$-wave, $d$-wave, and Sommerfeld-enhancement in the Coulomb limit.
We present the first effective $J$-factor analysis conducted for many of these dSphs under certain annihilation scenarios.
In particular, for Sagittarius~II, we perform the first $J$-factor analysis for any annihilation model.
Moreover, we are the first to our knowledge to calculate effective $J$-factors of any dSph for $d$-wave annihilation, as well as for $p$-wave annihilation without assuming a Maxwell-Boltzmann distribution.

We use a Navarro-Frenk-White (NFW) density profile for the dSph halos and assume that the DM velocity distribution is related to the density profile by the Eddington inversion formula~\cite{Widrow:2000fv}.
Under the approximation that a dSph spans a small angular size (which is well justified for all dSphs we consider), we employ previous work that has determined the effective $J$-factor in terms of the scale density, scale radius, and distance to the halo for all annihilation models we consider~\cite{Boddy:2018ike}.
We then estimate these parameters by fitting the associated velocity dispersion to stellar data and present results for the effective $J$-factors, integrated over various angular cones.

Finally, we use the \texttt{MADHAT} code~\cite{MADHAT} to perform a stacked analysis of Fermi-LAT gamma-ray data~\cite{Atwood:2009ez} for these targets.
We obtain bounds on the DM annihilation cross section for each of the annihilation models we consider.
Limits on Sommerfeld-enhanced annihilation and $p$-wave annihilation have been previously obtained using a smaller set of dSphs, with effective $J$-factors determined using different methodologies~\cite{Zhao2016,Boddy:2018qur}.

This paper is organized as follows.
In Sec.~\ref{sec:Jfactor}, we determine the effective $J$-factors for our set of 25 dSphs, describing our methodology in detail and comparing to previous results.
In Sec.~\ref{sec:bounds}, we use these effective $J$-factors, along with Fermi-LAT data, to set bounds on the DM annihilation cross section.
We conclude in Sec.~\ref{sec:conclusions}.

\section{Effective \texorpdfstring{$J$}{J}-factors}
\label{sec:Jfactor}

We express the DM annihilation cross section as $\sigma_A v = (\sigma_A v)_0 S(v/c)$, where $(\sigma_A v)_0$ is a quantity independent of the relative velocity $v$ of the annihilating particles.
The differential photon flux arising from DM annihilation in any astrophysical target is
\begin{equation}
  \frac{d^2 \Phi}{d\Omega\, dE_\gamma}
  = \frac{(\sigma_A v)_0}{8\pi m_X^2} J_S(\Omega) \frac{dN}{dE_\gamma} \ ,
\end{equation}
where $dN/dE_\gamma$ is the photon spectrum produced per annihilation and $m_X$ is the DM particle mass.
We have assumed that the DM particle is its own antiparticle.
The effective $J$-factor, $J_S(\Omega)$, encodes the information about the DM distribution in the target.
For a target with a central potential and DM particles on isotropic orbits, the DM velocity distribution $f(r,v_p)$ is simply a function of the distance from the center of the target and the velocity of the DM particle~\cite{Widrow:2000fv}.
With this simplification, the effective $J$-factor is
\begin{align}
  J_S (\theta) =& \int_0^\infty d\ell \int d^3 v_1 \int d^3 v_2 \,
  S(|\boldsymbol{v}_1-\boldsymbol{v}_2|/c) \nonumber\\
  & \times
  f\left[r(\ell, \theta),v_1 \right] \,
  f\left[r(\ell, \theta),v_2 \right] \ ,
  \label{eq:Js}
\end{align}
where $\ell$ is the distance along the line of sight and $\theta$ is the angle between the light of sight and the line to the center of the target.
The radial distance from the halo center can be recast via $r^2 (\ell, \theta) = \ell^2 + D^2 - 2\ell D \cos \theta$, where $D$ is the distance to the center of the target.

We consider DM annihilation models of the form $S(v/c) = (v/c)^n$, for integer $n$.
In particular, we focus on the following possible scenarios:
\begin{enumerate}[(i)]
\item{$n=-1$: Sommerfeld-enhanced annihilation in the Coulomb limit~\cite{ArkaniHamed:2008qn,Feng:2010zp}.
  If the annihilation proceeds through a heavy mediator, then $(\sigma v)_0 (2\pi \alpha_X) (v/c)^{-1}$, where $\alpha_X$ is the DM self coupling.
  We fix $\alpha_X = 1/2\pi$.}
\item{$n=0$: $s$-wave velocity-independent annihilation.
  This scenario is the one that is usually considered.}
\item{$n=2$: $p$-wave annihilation.
  This scenario is relevant if DM is a Majorana fermion, which annihilates to Standard Model fermion/antifermion pairs through an interaction that respects Minimal Flavor Violation.
  In this case, annihilation from an $s$-wave initial state is chirality-suppressed.
  As another example, this scenario is relevant if DM is a fermion (Dirac or Majorana) that annihilates through a scalar current coupling, regardless of the final state particles; in this case, the matrix element is only non-vanishing if the DM initial state is $p$-wave (see, for example, Ref.~\cite{Kumar:2013iva}).}
\item{$n=4$: $d$-wave annihilation.
  This scenario is relevant if DM is instead a real scalar~\cite{Giacchino:2013bta} that annihilates to Standard Model fermion/antifermion pairs through an interaction that respects Minimal Flavor Violation.
  In this case, annihilation from an $s$-wave state is chirality-suppressed, and the $p$-wave initial state is forbidden by symmetry of the wave function~\cite{Kumar:2013iva,Giacchino:2013bta}.}
\end{enumerate}

Following Ref.~\cite{Boddy:2019wfg}, we assume that the DM velocity distribution is a function of only two dimensionful parameters: the scale radius $r_s$ and the scale density $\rho_s$.
Furthermore, we take the limit $\theta_0 \ll 1$, where $\theta_0 \equiv r_s/D$ is the characteristic angular scale of the target.
Under these assumptions, the effective $J$-factor for a given annihilation model parameter $n$ may thus be written as
\begin{equation}
  J_{S(n)} (\tilde \theta) = 2 \rho_s^2 r_s
  \left(\frac{4\pi G_N \rho_s r_s^2}{c^2} \right)^{n/2}
  \tilde J_{S(n)} (\tilde \theta) \ ,
  \label{eq:JSn}
\end{equation}
where $\tilde{\theta} \equiv \theta / \theta_0$ and $\tilde J_S (\tilde \theta)$ is the scale-free angular distribution that depends only on $n$ and on the functional form of the velocity distribution, but not on the parameters $\rho_s$, $r_s$, or $D$.
Deviations from this result scale as $\mathcal{O}(\theta_0^2)$, which is negligible for the dSphs we consider.

Therefore, to determine the effective $J$-factor for any dSph, it is only necessary to know the halo parameters ($\rho_s$, $r_s$, and $D$) and scale-free angular distribution $\tilde J_{S(n)} (\tilde \theta)$.
The latter has been previously calculated for an NFW density profile $\rho (r) = \rho_s (r/r_s)^{-1} (1+r/r_s)^{-2}$ and for all values of $n$ discussed above~\cite{Boddy:2019wfg}.
We make the same assumptions as Ref.~\cite{Boddy:2019wfg} about the DM velocity distribution, which is related to the density profile through the Eddington inversion formula.
Using these results, we are able to determine the various effective $J$-factors for individual dSphs if we know $\rho_s$, $r_s$, and $D$.

In the following subsections, we describe our procedure of determining these parameters and present the resulting effective $J$-factors for specific dSphs.

\subsection{Halo parameters}

We use the halo parameter analysis, originally presented in Ref.~\cite{Pace:2018tin}, which calculated $J$-factors for 41 dSphs.
We consider the subset of 22 dSphs that have confidently measured velocity dispersions and are MW satellites.
The general methodology for determining halo parameters in dSphs is through a spherical Jeans analysis~\cite{Strigari2008ApJ...678..614S, Bonnivard2015MNRAS.453..849B, Geringer-Sameth2015ApJ...801...74G}.
The analysis involves solving the spherical Jeans equation (which relates the velocity dispersion, stellar anisotropy, and gravitational potential) for a set of halo parameters, projecting it into the line-of-sight direction, and comparing the line-of-sight dispersion to observed stellar kinematics.
For our spherical Jeans analysis, we assume a Plummer distribution for the stellar density~\cite{Plummer1911MNRAS..71..460P}, an NFW profile for the DM distribution~\cite{Navarro1997ApJ...490..493N}, and a constant stellar anisotropy.

This analysis includes a total of seven parameters: the three needed to find the effective $J$-factor ($\rho_s$, $r_s$, and $D$) and four others [average line-of-sight velocity, half-light radius ($r_p$), ellipticity ($\epsilon$), stellar anisotropy ($\beta$)].
The half-light radius, ellipticity, and distance all contain Gaussian priors based on literature measurements. For the halo parameters, we assume Jeffreys priors: $-2<\log_{10}{\left(r_s/{\rm kpc}\right)}<1$ and $4<\log_{10}{\left(\rho_s/(M_{\odot}\, {\rm kpc}^{-3})\right)}<14$.
The stellar anisotropy prior is uniform in a symmetrized version: $-0.95 < \tilde\beta <1$, where $\tilde\beta = \beta/(2-\beta)$.%
\footnote{Generally, $\beta$ ranges between $-\infty$ and $1$.  Negative and positive values correspond to tangential and radial orbits, respectively.  This alternate parameterization uniformly favors tangential and radial orbits.}
{To eliminate some unphysical points in the parameter space, we use the global density slope-anisotropy inequality,  $\gamma_\star(r) \geq 2 \beta(r)$, where $\gamma_\star$ is the log stellar density slope~\citep{Evans2009MNRAS.393L..50E, An2009ApJ...701.1500A, Ciotti2010MNRAS.408.1070C, Bonnivard2015MNRAS.446.3002B}. For a Plummer stellar profile and a constant stellar anisotropy, this constraint is $\beta<0$. }
We use an unbinned likelihood~\cite{Geringer-Sameth2015ApJ...801...74G} and determine the posterior distributions with a multimodal nested sampling algorithm~\cite{Feroz2008MNRAS.384..449F, Feroz2009MNRAS.398.1601F}.
We refer the reader to Ref.~\cite{Pace:2018tin} for more details.

We also apply the same analysis to three additional dSphs: Crater~II~\cite{Caldwell:2016hrl}, Hydrus~I~\cite{Koposov2018MNRAS.479.5343K}, and Sagittarius~II\footnote{We note that the identification of Sagittarius~II as a dSph versus a star cluster is not yet definite.
It has a very compact size and high luminosity compared to what is expected for a dSph~\cite{Laevens:2015kla,MutlaPakdil2018ApJ...863...25M}.
The velocity dispersion is resolved, but the mass-to-light ratio is much lower than other ultrafaint dSphs, and the metallicity dispersion is possibly resolved~\cite{Longeard2019arXiv190202780L}.
{Furthermore, there are other dwarf galaxies included in our sample whose nature is debated: Segue~1~\cite{Niederste-Ostholt2009MNRAS.398.1771N, Simon2011ApJ...733...46S} and Willman~1\cite{Siegel2008AJ....135.2084S,Willman2011AJ....142..128W}.} }.
The literature properties we use for our modeling are as follows: $D=117.5\pm1.1$ kpc, $r_p=31.1\pm2.5$ arcmin~\cite[Crater~II,][]{Torrealba2016MNRAS.459.2370T}; $D=27.6\pm0.5$ kpc, $r_p=7.42\pm0.58$ arcmin, $\epsilon=0.21\pm0.11$~\cite[Hydrus~I,][]{Koposov2018MNRAS.479.5343K}; and $D=73.1\pm0.9$ kpc, $r_p=1.7\pm0.05$ arcmin~\cite[Sagittarius~II,][]{Longeard2019arXiv190202780L}.
For Crater~II and Hydrus~I, we determine membership with a mixture model including a dSph and MW foreground component (see Ref.~\cite{Pace2020arXiv200209503P} for details on the mixture model).
We include {\it Gaia} DR2 proper motions, which helps identify dSph stars~\cite{GaiaBrown2018A&A...616A...1G, GaiaHelmi2018A&A...616A..12G}.
The $J$-factors of Hydrus~I and Crater~II have been presented before \cite{Caldwell:2016hrl,Koposov2018MNRAS.479.5343K}, and our results are comparable to previous measurements.
This is the first $J$-factor analysis of Sagittarius~II.

\subsection{Results}

\begin{figure*}[pt!]
  \centering
  \includegraphics[width=\textwidth]{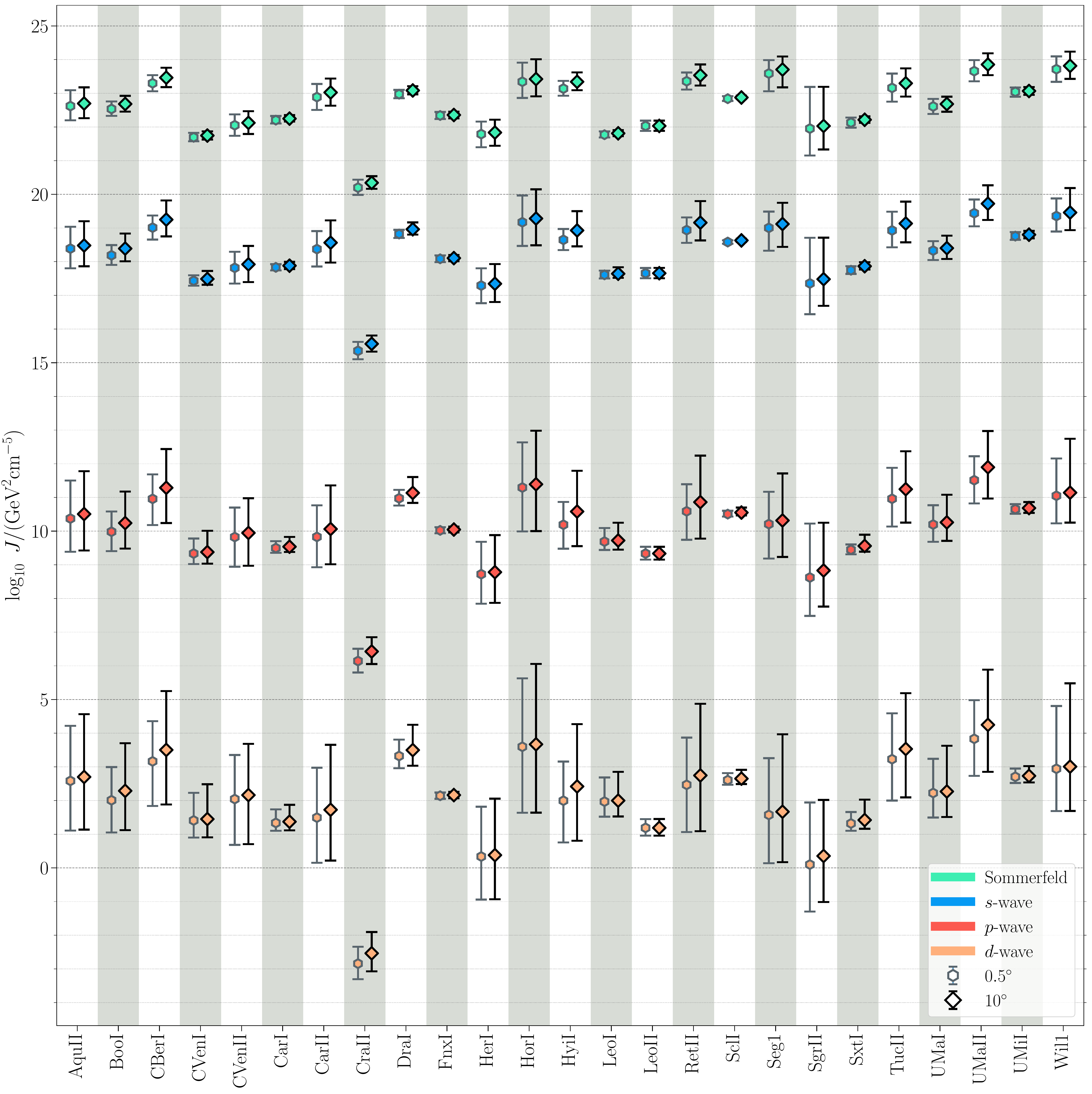}
  \caption{Median effective $J$-factors, integrated over cones with opening half-angles of $0.5^\circ$ (hexagons) and $10.0^\circ$ (diamonds), along with asymmetric 68\% containment bands.
  We show results for Sommerfeld-enhanced (green), $s$-wave (blue), $p$-wave (red), and $d$-wave (orange) DM annihilation.}
  \label{fig:JFactors}
\end{figure*}

Using the values of $r_s$, $\rho_s$, and $D$ from our nested sampling runs, we are able to produce the posterior distributions for the effective $J$-factor [from Eq.~\eqref{eq:JSn}], integrated over a given angular region of each dSph for different annihilation models.
In Fig.~\ref{fig:JFactors}, we show the effective $J$-factors for our set of 25 dSphs, integrated over cones with opening half-angles of $0.5^\circ$ and $10.0^\circ$, for the scenarios of $s$-wave, $p$-wave, $d$-wave, and Sommerfeld-enhanced annihilation.
In each case, we plot the median $J$-factor, along with $68\%$ containment bands.
Note that, in general, there is very little difference between the effective $J$-factor integrated over cones with half-angles of $0.5^\circ$ and $10.0^\circ$, indicating that in most cases the $0.5^\circ$ cone encompasses the dSph almost entirely.

Table~\ref{tab:j_full} lists the median integrated effective $J$-factors with their $1\sigma$ uncertainties for each of our annihilation scenarios.
We provide effective $J$-factors integrated over cones with half-angles of $0.1^\circ$, $0.2^\circ$, $0.5^\circ$, and $10^\circ$\footnote{The posterior distributions are available at the following website: \url{https://github.com/apace7/J-Factor-Scaling} }.

\subsection{Comparison with other approaches}

\begin{figure*}[h]
    \centering
    \includegraphics[width=\textwidth]{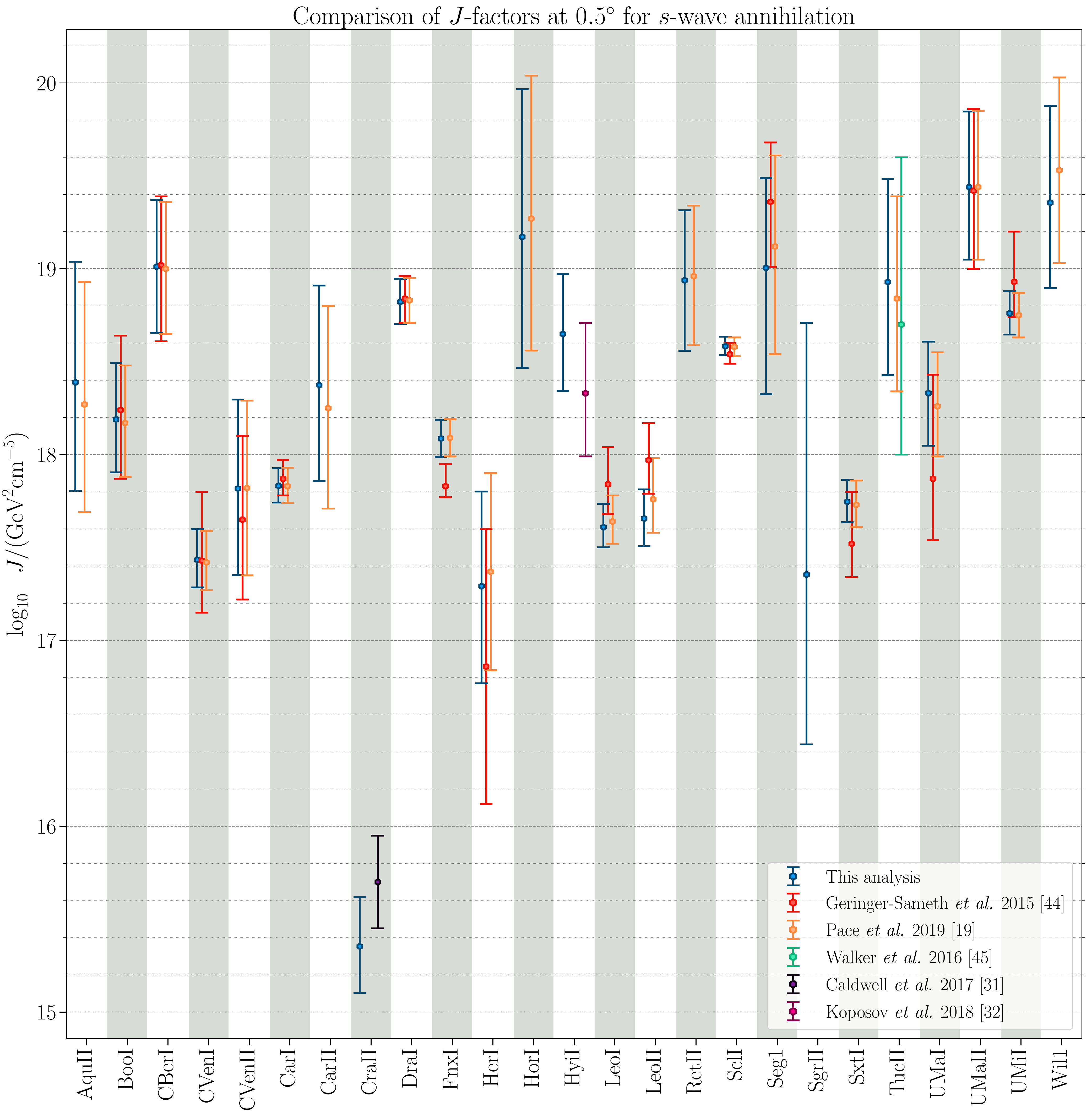}
    \caption{$J$-factors for $s$-wave DM annihilation integrated over a $0.5^\circ$ cone from this analysis and others.}
    \label{fig:comp_plot_s_half}
\end{figure*}

\begin{figure*}[h]
    \centering
    \includegraphics[width=\textwidth]{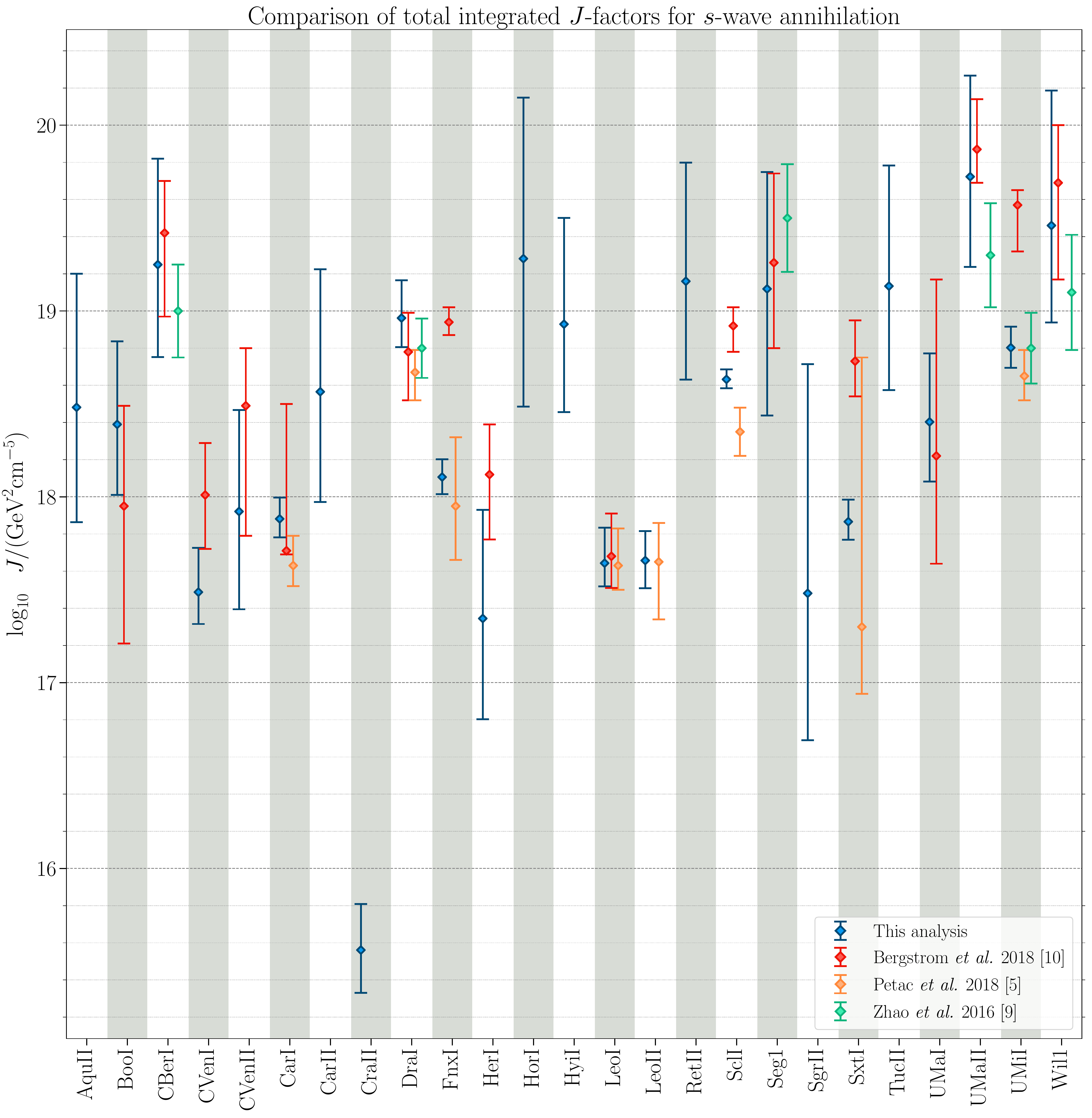}
    \caption{$J$-factors for $s$-wave DM annihilation integrated over a $10^\circ$ cone from this analysis and total $J$-factors from other analyses.}
    \label{fig:comp_plot_s_tot}
\end{figure*}

\begin{figure*}[h]
    \centering
    \includegraphics[width=\textwidth]{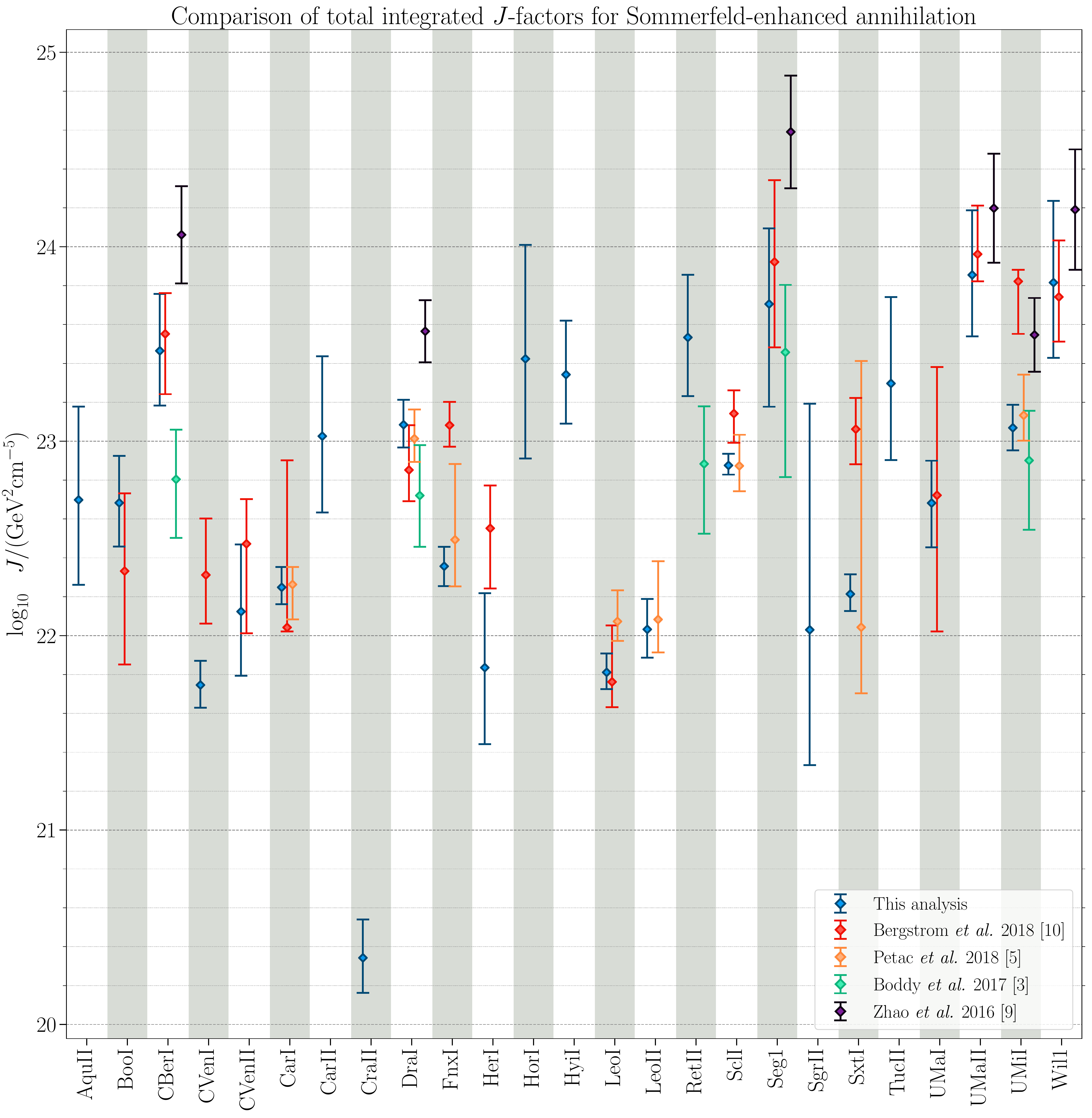}
    \caption{Effective $J$-factors for Sommerfeld-enhanced DM annihilation integrated over a $10^\circ$ cone from this analysis and total $J$-factors from other analyses.
      The total effective $J$-factors from other analyses have been rescaled to $\alpha_X = 1/2\pi$ for comparison.}
    \label{fig:comp_plot_som}
\end{figure*}

\begin{figure*}[h]
    \centering
    \includegraphics[width=\textwidth]{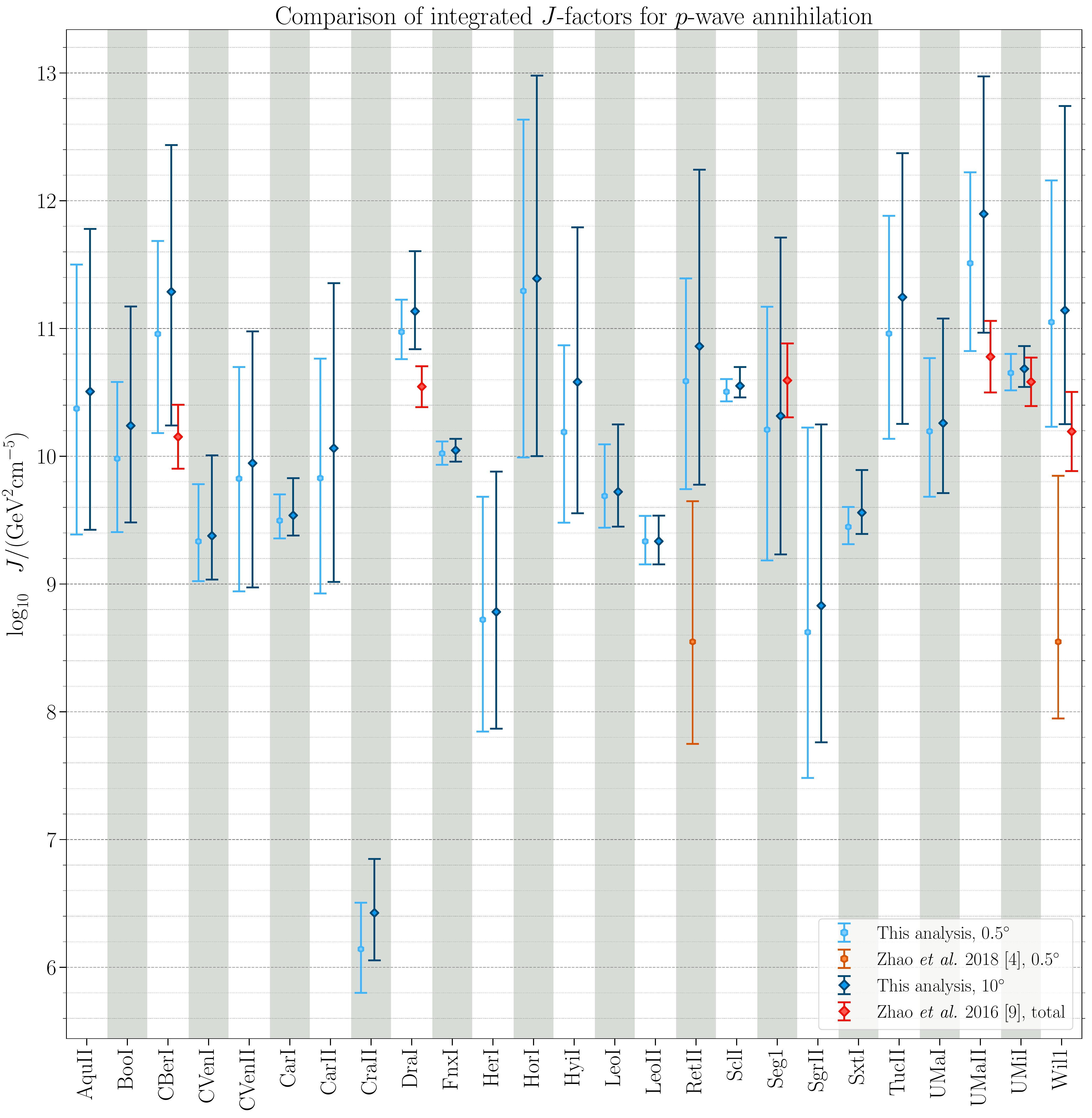}
    \caption{Effective $J$-factors for $p$-wave DM annihilation, integrated over a cone of $0.5^\circ$ (hexagons) or $10.0^\circ$ (diamonds), or the total integrated effective $J$-factor (diamonds), as indicated in the legend.}
    \label{fig:comp_plot_p}
\end{figure*}

We compare our results to other results found in the literature.
In Fig.~\ref{fig:comp_plot_s_half}, we compare our results for the $s$-wave $J$-factor (integrated over a $0.5^\circ$ cone) to those found in Refs.~\cite{Geringer-Sameth:2014qqa,Pace:2018tin,Walker:2016adk,Caldwell:2016hrl,Koposov2018MNRAS.479.5343K}.
In Fig.~\ref{fig:comp_plot_s_tot}, we compare our results for the $s$-wave $J$-factor (integrated over a $10^\circ$ cone) to the total integrated $J$-factors found in Refs.~\cite{Bergstrom2018,Petac:2018gue,Zhao2016}.
In Fig.~\ref{fig:comp_plot_som}, we compare our results for the Sommerfeld-enhanced effective $J$-factor (integrated over a $10^\circ$ cone) to the total Sommerfeld-enhanced effective $J$-factors found in Refs.~\cite{Bergstrom2018,Petac:2018gue,Boddy:2017vpe,Zhao2016}.%
\footnote{All of the effective $J$-factors in these other works have been rescaled to $\alpha_X = 1/2\pi$ for direct comparison with our calculations.}
In Fig.~\ref{fig:comp_plot_p}, we compare our results for the $p$-wave effective $J$-factor (integrated over a $0.5^\circ$ cone) to those in Ref.~\cite{Zhao2017} and for the $p$-wave effective $J$-factor (integrated over a $10^\circ$ cone) to the total $p$-wave effective $J$-factor found in Ref.~\cite{Zhao2016}.

We note here one detail regarding the comparison of our results to those in Ref.~\cite{Zhao2016} for the case of $p$-wave or Sommerfeld-enhanced annihilation.
In Ref.~\cite{Zhao2016}, the DM velocity distribution is assumed to be Maxwell-Boltzmann, with a velocity dispersion that is independent of position.
In this case, the velocity and position integrals in Eq.~\eqref{eq:Js} factorize, and the total effective $J$-factor can be written as the product of the total $s$-wave $J$-factor and a velocity integral that depends only on the assumed velocity dispersion.
In Ref.~\cite{Zhao2016}, this integral is absorbed into the definition of the thermally averaged annihilation cross section.
For the purposes of comparison, we have rescaled the $s$-wave $J$-factors reported in Ref.~\cite{Zhao2016} by the appropriate integrals, in order to obtain their total effective $J$-factors.
For the case of $p$-wave [or Sommerfeld-enhanced] annihilation, the rescaling factor is $6 (v_0/c)^2$ [or $\pi^{-1/2} (v_0/c)^{-1}$], where the values of the velocity dispersion $v_0$ are also taken from Ref.~\cite{Zhao2016}.

\section{Constraints on Dark Matter Annihilation}
\label{sec:bounds}

Having determined the effective $J$-factors for our set of dSphs under different DM annihilation models, we now constrain DM annihilation to a variety of Standard Model final states by performing a stacked dSph analysis with Fermi-LAT gamma-ray data.
We use the \texttt{MADHAT 1.0} software package~\cite{MADHAT}, which is based on the model-independent analysis framework described in Ref.~\cite{Boddy:2018qur}.
\texttt{MADHAT} uses Fermi-LAT Pass 8R3 data~\cite{Bruel:2018lac}, collected over a time frame of nearly 11 years, and incorporates gamma rays only in the energy range of 1-100~GeV, across which the Fermi-LAT effective area is treated as approximately constant.
This process makes it possible to apply a stacked analysis to any particle physics model in this energy range without having to process Fermi-LAT data for a given analysis to account for the energy dependence of the detector.
We simply need the gamma-ray spectrum $dN/dE_\gamma$ for a specific annihilation channel to produce bounds with \texttt{MADHAT}, and we obtain relevant spectra from the tools described in Ref.~\cite{Cirelli:2010xx}.

\begin{figure*}[ht!]
  \centering
  \includegraphics[width=\textwidth]{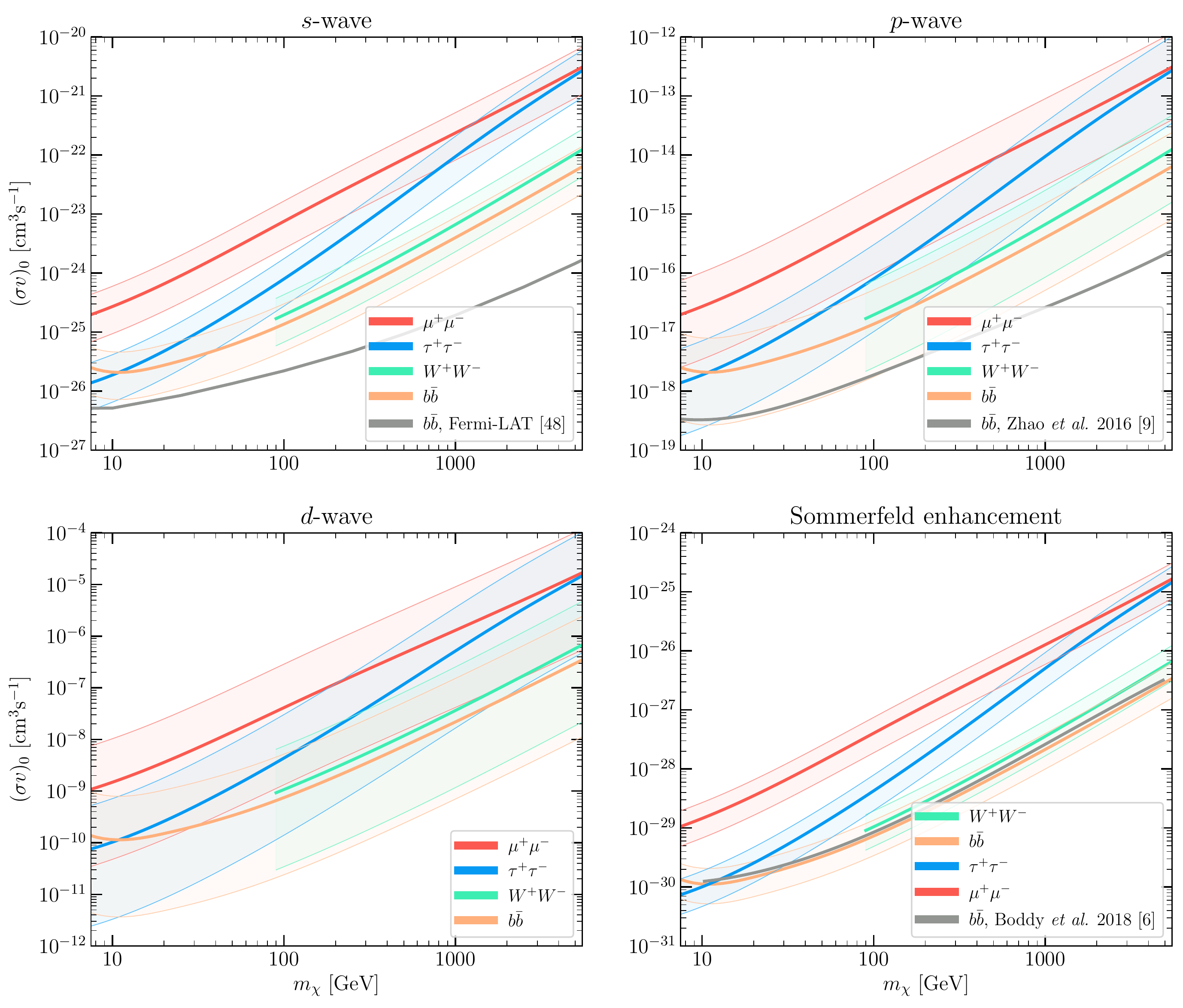}
  \caption{Exclusion limits at 95\% C.L.~for $s$-wave (upper left), $p$-wave (upper right), $d$-wave (lower left) and Sommerfeld-enhanced (in the Coulomb limit, lower right) DM annihilation.
    We consider the following annihilation channels: $\bar b b$ (orange), $\bar \tau \tau$ (blue), $\bar \mu \mu$ (red), and $W^+ W^-$ (green).
    Each central solid line is obtained using the median effective $J$-factors for all 25 dSphs considered, while the colored bands indicate the limits obtained by varying all effective $J$-factors either up or down by their $1\sigma$ uncertainty.
    The solid gray lines reproduce the exclusion limits for annihilation to the $\bar b b$ channel at 95\% C.L.~found in  Ref.~\cite{Ackermann:2015zua} ($s$-wave), Ref.~\cite{Zhao2016} [$p$-wave, expressed as a bound on $(\sigma v)_0$], and Ref.~\cite{Boddy:2018qur} (Sommerfeld-enhanced, rescaled to $\alpha_X =  1/2\pi$).}
  \label{fig:Constraints}
\end{figure*}

In Fig.~\ref{fig:Constraints}, we plot constraints on $s$-wave, $p$-wave, $d$-wave, and Sommerfeld-enhanced DM annihilation to $\bar{b} b$, $\bar{\tau} \tau$, $\bar{\mu} \mu$, and $W^+ W^-$.
For each channel, the solid line is the 95\% C.L.~bound derived from an analysis of all 25 dSphs, setting the effective $J$-factors (integrated over a cone with an opening half-angle of $0.5^\circ$) to their median values, while the uncertainty band arises from adjusting their values upward or downward by their $1\sigma$ systematic uncertainties.%
\footnote{We also obtained 95\% C.L.~bounds by stacking only dSphs with integrated effective $J$-factors that are at least 15\% of the largest integrated effective $J$-factor. Using this subset of dSphs did not affect the limits in any significant way.}

For comparison, we also plot 95\% C.L.~bounds on annihilation to $\bar{b} b$ from other analyses, represented by the solid gray lines in Fig.~\ref{fig:Constraints}.
For the $s$-wave scenario, we show the bounds from the Fermi collaboration analysis~\cite{Ackermann:2015zua}.
These constraints are stronger than ours, but the two results are in agreement within the level of their uncertainty bands (the 95\% containment bands from Ref.~\cite{Ackermann:2015zua} are not shown) for $m_X \lesssim 100$~GeV.
At higher DM masses, our $s$-wave bounds are considerably weaker than those found in Ref.~\cite{Ackermann:2015zua}.
This discrepancy is likely due to our analysis using \texttt{MADHAT}, which limits the photon energy range to $< 100$~GeV in order to achieve a model-independent analysis framework, while the analysis in Ref.~\cite{Ackermann:2015zua} uses photons up to an energy of $500$~GeV.

We also show the bounds on $p$-wave annihilation found in Ref.~\cite{Zhao2016}, recast as a bound on $(\sigma v)_0$.
Finally, to provide a comparison for Sommerfeld-enhanced annihilation in the Coulomb limit, we obtain bounds using the effective $J$-factors found in Ref.~\cite{Boddy:2017vpe}, rescaled to $\alpha_X = 1/2\pi$.
In both cases, these bounds lie within our $1\sigma$ systematic uncertainty band for $m_\chi < {\cal O}(100)$~GeV.

As bounds on $d$-wave annihilation have not previously been determined, we comment on the applicability of our $d$-wave constraints.
If DM is a real scalar that annihilates to a fermion/antifermion pair ($\phi \phi \rightarrow \bar{f} f$) through an interaction respecting Minimal Flavor Violation, then annihilation from a $p$-wave initial state is exactly forbidden, while annihilation from an $s$-wave initial state is chirality-suppressed by a factor $\sim (m_f / m_\phi)^2$.
For DM of mass $m_X \sim {\cal O}(10)$ TeV annihilating to muons, the suppression factor is $\sim {\cal O}(10^{-10})$, which is still larger than the $v^4$-suppression factor associated with $d$-wave DM annihilation in a dSph.
As such, absent fine-tuning, we expect $d$-wave annihilation to dominate $s$-wave annihilation for DM much heavier than ${\cal O}(10)$ TeV, in which case the Fermi-LAT would not be the ideal instrument to set constraints.
Note that these considerations are not necessarily relevant for constraints on $p$-wave annihilation, since there are scenarios in which DM annihilation from an $s$-wave initial state is effectively forbidden, while DM annihilation from a $p$-wave state is necessarily dominant.

\section{Conclusions}
\label{sec:conclusions}

There are well-motivated DM models that produce annihilation cross sections with power-law scalings of the relative velocity.
For scenarios beyond velocity-independent $s$-wave annihilation, standard calculations of the astrophysical $J$-factor for indirect detection must be modified to account for any velocity dependence.
Under various simplifying assumptions, we can infer the DM halo velocity distribution from the density profile using the Eddington inversion formula.
With the velocity distribution at hand, we incorporate the velocity dependence of the annihilation cross section into the calculation of the $J$-factor to produce an effective $J$-factor.

We have determined these effective $J$-factors for 25 dwarf spheroidal galaxies, with assumed NFW halo profiles, for DM annihilation that is $s$-wave, $p$-wave, $d$-wave, or Sommerfeld-enhanced in the Coulomb limit.
We present the first analysis that we are aware of for several dSphs under certain annihilation models.
In particular, we perform the first analysis for Sagittarius~II for any annihilation model.
Changing the assumed particle physics model for DM annihilation can change the effective $J$-factor by several orders of magnitude.

We have used these effective $J$-factors and the \texttt{MADHAT} software package to determine bounds on DM annihilation in each of these scenarios with a stacked analysis of Fermi-LAT gamma-ray data from dSphs.
The limits on $s$-wave annihilation are consistent with those found previously in the literature.

Although we have assumed the dSphs have an NFW density profile, similar methods can be used for other profiles, such as the generalized NFW, Burkert, Einasto, etc.
It would be interesting to see how the choice of a different profile affects the effective $J$-factors for non-$s$-wave models of DM annihilation.
{Another avenue for improvement in our analysis is to lift the assumption of spherical symmetry.  This has been explored in the $s$-wave $J$-factor case but has not yet been extended to velocity dependent models~\citep{Evans2016PhRvD..93j3512E, Sanders2016PhRvD..94f3521S, Hayashi2016MNRAS.461.2914H, Klop2017PhRvD..95l3012K}. }
We leave such a study to future work.

Observations of systems other than dSphs may provide competitive or stronger limits on velocity-dependent annihilation.
For instance, typical DM particle velocities can be quite small in the early Universe, and thus strong constraints on Sommerfeld-enhanced DM annihilation arise from observations of the cosmic microwave background and measurements of light element abundances~\cite{Finkbeiner:2010sm,kohri2011}.
How competitive the cosmological constraints are compared to dSphs, however, is model-dependent: the velocity behavior of the Sommerfeld enhancement depends on the mass of the particle mediating the dark matter self-interaction.

While limits on $p$-wave and $d$-wave annihilation may be stronger from systems with larger characteristic velocities, such as clusters or the Galactic Center, the limits we have derived are robust, because the DM distributions in dSphs are directly extracted from the data.
Since baryons contribute a non-negligible amount to the potential of clusters and MW-like galaxies in the regions where the DM annihilation signal arises, they represent an important systematic uncertainty that must be dealt with in these systems.
Previous studies of $p$-wave DM annihilation in the Galactic Center considered the increase in density and characteristic velocities near the central black hole~\cite{Shelton:2015aqa,Sandick:2016zeg,Johnson:2019hsm} and used these to argue for stronger constraints on $p$-wave models.
There are also constraints on $p$-wave annihilation from the epoch of reionization, for which the results depend on assumptions of the reionization history of the intergalactic medium and the structure formation prescription used to determine annihilation boost factors~\cite{Liu:2016cnk}.
Future observatories, such as CTA~\cite{Acharya:2017ttl}, will target the Galactic Center in particular, and there are a variety of upcoming and proposed instruments that will improve our understanding of reionization.
It is thus important to compare the ultimate sensitivity that these types of observations can achieve in comparison to dSphs.

We expect the discovery of many new dSphs from current instruments, such as DECam \cite[e.g., ][]{Bechtol2015ApJ...807...50B, Drlica-Wagner2015ApJ...813..109D} and Hyper-Surprime Cam \cite{Homma2018PASJ...70S..18H, Homma2019PASJ..tmp...91H}, as well as from future observatories, such as the LSST \citep{ivezic2008lsst, Hargis2014ApJ...795L..13H, Newton2018MNRAS.479.2853N}.
If nearby dSphs with large effective $J$-factors are found, observational sensitivity to DM annihilation could improve significantly.
It is standard practice to estimate $s$-wave $J$-factors for dSphs and dSph-candidates, but we have demonstrated that it is just as straightforward to estimate the effective $J$-factors relevant for velocity-dependent DM annihilation.
Since the fundamental nature of DM interactions is still mysterious, it is important to use data to search for and constrain a variety of DM annihilation models.


\acknowledgments
We are grateful to Stephen Hill for useful discussions.
JK is grateful to the organizers of the Mitchell Conference on Collider, Dark Matter and Neutrino Physics 2019, where work on this project began, for their hospitality.
The work of JK is supported in part by DOE Grant de-sc0010504.
The work of ABP is supported by NSF grant AST-1813881.
The work of LES is supported by DOE Grant de-sc0010813.

This work has made use of data from the European Space Agency (ESA) mission {\it Gaia} (\url{https://www.cosmos.esa.int/gaia}), processed by the {\it Gaia} Data Processing and Analysis Consortium (DPAC, \url{https://www.cosmos.esa.int/web/gaia/dpac/consortium}).
Funding for the DPAC has been provided by national institutions, in particular the institutions participating in the {\it Gaia} Multilateral Agreement.


\bibliography{dSph_Jfactors}


\begin{turnpage}
\begin{table*}[h]
    \input{full_table.txt}
    \caption{Integrated $J$-factors presented as $\log_{10}(J/\mathrm{GeV}^2\mathrm{cm}^{-5})$ for $s$-wave, $p$-wave, $d$-wave, and Sommerfeld-enhanced annihilation for cones with opening half-angles of 0.1$^\circ$, 0.2$^\circ$, 0.5$^\circ$, and 10$^\circ$.
      The posterior distributions are available at the following website: \url{https://github.com/apace7/J-Factor-Scaling}.}
    \label{tab:j_full}
\end{table*}
\end{turnpage}

\end{document}

%% file: full_table.txt
\centering\footnotesize\hspace*{-2cm}\begin{tabular}{| l | c c c c | c c c c | c c c c | c c c c |}\cline{2-17}\multicolumn{1}{c}{} & \multicolumn{4}{|c|}{$s$-wave} & \multicolumn{4}{|c|}{$p$-wave} & \multicolumn{4}{|c|}{$d$-wave} & \multicolumn{4}{|c|}{Sommerfeld}\\\hline Galaxy & $0.1^\circ$ & $0.2^\circ$ & $0.5^\circ$ & $10^\circ$& $0.1^\circ$ & $0.2^\circ$ & $0.5^\circ$ & $10^\circ$& $0.1^\circ$ & $0.2^\circ$ & $0.5^\circ$ & $10^\circ$& $0.1^\circ$ & $0.2^\circ$ & $0.5^\circ$ & $10^\circ$\\\hline 
Aquarius II & $18.12^{+0.62}_{-0.56}$ & $18.26^{+0.63}_{-0.57}$ & $18.39^{+0.65}_{-0.58}$ & $18.48^{+0.72}_{-0.62}$ & $9.99^{+0.98}_{-0.87}$ & $10.19^{+1.03}_{-0.90}$ & $10.37^{+1.13}_{-0.99}$ & $10.51^{+1.27}_{-1.08}$ & $2.11^{+1.44}_{-1.25}$ & $2.37^{+1.52}_{-1.34}$ & $2.58^{+1.63}_{-1.47}$ & $2.70^{+1.86}_{-1.56}$ & $22.45^{+0.48}_{-0.43}$ & $22.54^{+0.48}_{-0.42}$ & $22.62^{+0.47}_{-0.42}$ & $22.70^{+0.48}_{-0.44}$\\
Bo\"{o}tes I & $17.79^{+0.30}_{-0.28}$ & $17.99^{+0.29}_{-0.27}$ & $18.19^{+0.30}_{-0.28}$ & $18.39^{+0.45}_{-0.38}$ & $9.38^{+0.45}_{-0.42}$ & $9.68^{+0.49}_{-0.46}$ & $9.98^{+0.60}_{-0.58}$ & $10.24^{+0.93}_{-0.76}$ & $1.30^{+0.74}_{-0.68}$ & $1.65^{+0.83}_{-0.78}$ & $2.01^{+0.99}_{-0.96}$ & $2.29^{+1.42}_{-1.17}$ & $22.28^{+0.26}_{-0.25}$ & $22.41^{+0.24}_{-0.23}$ & $22.54^{+0.22}_{-0.21}$ & $22.68^{+0.24}_{-0.23}$\\
Canes Venatici I & $17.19^{+0.18}_{-0.17}$ & $17.33^{+0.15}_{-0.14}$ & $17.43^{+0.16}_{-0.15}$ & $17.49^{+0.24}_{-0.17}$ & $9.01^{+0.23}_{-0.21}$ & $9.20^{+0.30}_{-0.25}$ & $9.33^{+0.45}_{-0.31}$ & $9.38^{+0.63}_{-0.34}$ & $1.07^{+0.45}_{-0.35}$ & $1.27^{+0.60}_{-0.43}$ & $1.41^{+0.82}_{-0.51}$ & $1.45^{+1.04}_{-0.54}$ & $21.53^{+0.21}_{-0.18}$ & $21.62^{+0.17}_{-0.15}$ & $21.69^{+0.13}_{-0.12}$ & $21.75^{+0.12}_{-0.12}$\\
Canes Venatici II & $17.53^{+0.42}_{-0.41}$ & $17.68^{+0.44}_{-0.43}$ & $17.82^{+0.48}_{-0.47}$ & $17.92^{+0.55}_{-0.53}$ & $9.39^{+0.71}_{-0.70}$ & $9.62^{+0.76}_{-0.77}$ & $9.82^{+0.87}_{-0.88}$ & $9.95^{+1.03}_{-0.97}$ & $1.54^{+1.08}_{-1.09}$ & $1.81^{+1.17}_{-1.22}$ & $2.04^{+1.31}_{-1.36}$ & $2.16^{+1.52}_{-1.46}$ & $21.87^{+0.33}_{-0.31}$ & $21.96^{+0.32}_{-0.31}$ & $22.05^{+0.32}_{-0.31}$ & $22.12^{+0.35}_{-0.33}$\\
Carina I & $17.57^{+0.17}_{-0.14}$ & $17.72^{+0.13}_{-0.11}$ & $17.83^{+0.09}_{-0.09}$ & $17.88^{+0.11}_{-0.10}$ & $9.16^{+0.13}_{-0.12}$ & $9.37^{+0.13}_{-0.12}$ & $9.50^{+0.21}_{-0.14}$ & $9.54^{+0.29}_{-0.16}$ & $0.98^{+0.18}_{-0.15}$ & $1.20^{+0.27}_{-0.18}$ & $1.34^{+0.40}_{-0.23}$ & $1.37^{+0.50}_{-0.26}$ & $22.04^{+0.18}_{-0.16}$ & $22.13^{+0.15}_{-0.13}$ & $22.20^{+0.12}_{-0.10}$ & $22.25^{+0.11}_{-0.09}$\\
Carina II & $18.02^{+0.53}_{-0.51}$ & $18.19^{+0.52}_{-0.50}$ & $18.37^{+0.54}_{-0.52}$ & $18.57^{+0.66}_{-0.59}$ & $9.29^{+0.81}_{-0.77}$ & $9.57^{+0.83}_{-0.82}$ & $9.83^{+0.94}_{-0.90}$ & $10.06^{+1.29}_{-1.05}$ & $0.91^{+1.18}_{-1.15}$ & $1.21^{+1.30}_{-1.23}$ & $1.49^{+1.48}_{-1.34}$ & $1.73^{+1.93}_{-1.51}$ & $22.66^{+0.43}_{-0.41}$ & $22.78^{+0.41}_{-0.39}$ & $22.89^{+0.39}_{-0.38}$ & $23.03^{+0.41}_{-0.39}$\\
Coma Berenices I & $18.61^{+0.31}_{-0.32}$ & $18.81^{+0.32}_{-0.32}$ & $19.01^{+0.36}_{-0.36}$ & $19.25^{+0.57}_{-0.50}$ & $10.32^{+0.54}_{-0.53}$ & $10.63^{+0.61}_{-0.61}$ & $10.96^{+0.73}_{-0.78}$ & $11.29^{+1.15}_{-1.05}$ & $2.40^{+0.94}_{-0.92}$ & $2.77^{+1.03}_{-1.07}$ & $3.16^{+1.19}_{-1.32}$ & $3.50^{+1.75}_{-1.62}$ & $23.03^{+0.27}_{-0.26}$ & $23.16^{+0.25}_{-0.25}$ & $23.30^{+0.24}_{-0.24}$ & $23.46^{+0.29}_{-0.28}$\\
Crater II & $14.91^{+0.33}_{-0.29}$ & $15.13^{+0.30}_{-0.28}$ & $15.35^{+0.27}_{-0.25}$ & $15.56^{+0.25}_{-0.23}$ & $5.48^{+0.41}_{-0.38}$ & $5.81^{+0.39}_{-0.35}$ & $6.14^{+0.36}_{-0.34}$ & $6.43^{+0.42}_{-0.37}$ & $-3.60^{+0.49}_{-0.46}$ & $-3.22^{+0.48}_{-0.46}$ & $-2.84^{+0.50}_{-0.46}$ & $-2.54^{+0.63}_{-0.54}$ & $19.92^{+0.28}_{-0.25}$ & $20.06^{+0.26}_{-0.24}$ & $20.20^{+0.24}_{-0.21}$ & $20.34^{+0.20}_{-0.18}$\\
Draco I & $18.43^{+0.16}_{-0.14}$ & $18.63^{+0.13}_{-0.13}$ & $18.82^{+0.12}_{-0.12}$ & $18.96^{+0.20}_{-0.16}$ & $10.41^{+0.15}_{-0.14}$ & $10.70^{+0.17}_{-0.15}$ & $10.97^{+0.25}_{-0.21}$ & $11.13^{+0.47}_{-0.30}$ & $2.68^{+0.27}_{-0.21}$ & $3.01^{+0.35}_{-0.26}$ & $3.32^{+0.49}_{-0.36}$ & $3.50^{+0.75}_{-0.46}$ & $22.72^{+0.18}_{-0.15}$ & $22.85^{+0.15}_{-0.14}$ & $22.97^{+0.13}_{-0.12}$ & $23.08^{+0.13}_{-0.12}$\\
Fornax I & $17.91^{+0.12}_{-0.13}$ & $18.02^{+0.11}_{-0.11}$ & $18.09^{+0.10}_{-0.10}$ & $18.11^{+0.10}_{-0.09}$ & $9.78^{+0.12}_{-0.13}$ & $9.93^{+0.10}_{-0.10}$ & $10.02^{+0.09}_{-0.09}$ & $10.05^{+0.09}_{-0.09}$ & $1.86^{+0.12}_{-0.12}$ & $2.04^{+0.10}_{-0.10}$ & $2.14^{+0.10}_{-0.09}$ & $2.16^{+0.10}_{-0.10}$ & $22.22^{+0.12}_{-0.13}$ & $22.30^{+0.11}_{-0.12}$ & $22.34^{+0.10}_{-0.11}$ & $22.36^{+0.10}_{-0.10}$\\
Hercules I & $17.09^{+0.49}_{-0.52}$ & $17.20^{+0.49}_{-0.51}$ & $17.29^{+0.51}_{-0.52}$ & $17.35^{+0.58}_{-0.54}$ & $8.46^{+0.77}_{-0.78}$ & $8.61^{+0.84}_{-0.82}$ & $8.72^{+0.96}_{-0.88}$ & $8.78^{+1.10}_{-0.91}$ & $0.05^{+1.18}_{-1.12}$ & $0.21^{+1.32}_{-1.20}$ & $0.34^{+1.48}_{-1.28}$ & $0.38^{+1.68}_{-1.31}$ & $21.65^{+0.39}_{-0.41}$ & $21.72^{+0.38}_{-0.39}$ & $21.79^{+0.37}_{-0.39}$ & $21.84^{+0.38}_{-0.39}$\\
Horologium I & $18.92^{+0.75}_{-0.62}$ & $19.05^{+0.75}_{-0.65}$ & $19.17^{+0.80}_{-0.70}$ & $19.28^{+0.87}_{-0.80}$ & $10.93^{+1.19}_{-1.07}$ & $11.12^{+1.25}_{-1.18}$ & $11.29^{+1.34}_{-1.30}$ & $11.39^{+1.59}_{-1.39}$ & $3.23^{+1.71}_{-1.67}$ & $3.42^{+1.84}_{-1.80}$ & $3.60^{+2.03}_{-1.96}$ & $3.67^{+2.39}_{-2.03}$ & $23.20^{+0.58}_{-0.48}$ & $23.27^{+0.58}_{-0.47}$ & $23.34^{+0.57}_{-0.48}$ & $23.42^{+0.59}_{-0.51}$\\
Hydrus I & $18.21^{+0.31}_{-0.29}$ & $18.42^{+0.30}_{-0.28}$ & $18.65^{+0.32}_{-0.31}$ & $18.93^{+0.57}_{-0.47}$ & $9.51^{+0.49}_{-0.46}$ & $9.83^{+0.55}_{-0.53}$ & $10.19^{+0.68}_{-0.71}$ & $10.58^{+1.21}_{-1.03}$ & $1.20^{+0.90}_{-0.84}$ & $1.58^{+1.00}_{-0.99}$ & $2.00^{+1.16}_{-1.24}$ & $2.42^{+1.85}_{-1.61}$ & $22.86^{+0.28}_{-0.26}$ & $23.00^{+0.25}_{-0.24}$ & $23.14^{+0.23}_{-0.21}$ & $23.34^{+0.28}_{-0.25}$\\
Leo I & $17.38^{+0.13}_{-0.11}$ & $17.52^{+0.10}_{-0.10}$ & $17.61^{+0.13}_{-0.11}$ & $17.64^{+0.19}_{-0.12}$ & $9.41^{+0.16}_{-0.14}$ & $9.58^{+0.25}_{-0.19}$ & $9.69^{+0.40}_{-0.25}$ & $9.72^{+0.53}_{-0.27}$ & $1.65^{+0.40}_{-0.27}$ & $1.84^{+0.54}_{-0.36}$ & $1.97^{+0.72}_{-0.45}$ & $2.00^{+0.85}_{-0.47}$ & $21.62^{+0.17}_{-0.13}$ & $21.70^{+0.13}_{-0.11}$ & $21.77^{+0.10}_{-0.08}$ & $21.81^{+0.10}_{-0.09}$\\
Leo II & $17.63^{+0.16}_{-0.16}$ & $17.65^{+0.15}_{-0.15}$ & $17.66^{+0.16}_{-0.15}$ & $17.66^{+0.16}_{-0.15}$ & $9.30^{+0.20}_{-0.18}$ & $9.33^{+0.20}_{-0.18}$ & $9.33^{+0.20}_{-0.18}$ & $9.33^{+0.20}_{-0.18}$ & $1.16^{+0.24}_{-0.23}$ & $1.18^{+0.26}_{-0.23}$ & $1.19^{+0.26}_{-0.23}$ & $1.19^{+0.26}_{-0.23}$ & $22.02^{+0.16}_{-0.17}$ & $22.03^{+0.16}_{-0.15}$ & $22.03^{+0.16}_{-0.15}$ & $22.03^{+0.16}_{-0.15}$\\
Reticulum II & $18.54^{+0.34}_{-0.32}$ & $18.73^{+0.34}_{-0.32}$ & $18.94^{+0.38}_{-0.38}$ & $19.16^{+0.64}_{-0.53}$ & $9.99^{+0.56}_{-0.56}$ & $10.28^{+0.65}_{-0.66}$ & $10.59^{+0.80}_{-0.85}$ & $10.86^{+1.38}_{-1.08}$ & $1.78^{+1.04}_{-0.99}$ & $2.12^{+1.17}_{-1.16}$ & $2.47^{+1.40}_{-1.40}$ & $2.75^{+2.13}_{-1.66}$ & $23.10^{+0.30}_{-0.28}$ & $23.23^{+0.28}_{-0.26}$ & $23.36^{+0.26}_{-0.25}$ & $23.53^{+0.32}_{-0.30}$\\
Sagittarius II & $17.07^{+1.60}_{-1.09}$ & $17.22^{+1.47}_{-1.02}$ & $17.35^{+1.36}_{-0.91}$ & $17.48^{+1.23}_{-0.79}$ & $8.20^{+1.95}_{-1.39}$ & $8.42^{+1.77}_{-1.27}$ & $8.62^{+1.60}_{-1.14}$ & $8.83^{+1.42}_{-1.07}$ & $-0.39^{+2.20}_{-1.61}$ & $-0.13^{+2.02}_{-1.50}$ & $0.10^{+1.84}_{-1.40}$ & $0.35^{+1.66}_{-1.37}$ & $21.77^{+1.39}_{-0.92}$ & $21.87^{+1.31}_{-0.87}$ & $21.95^{+1.24}_{-0.80}$ & $22.03^{+1.16}_{-0.70}$\\
Sculptor I & $18.33^{+0.11}_{-0.10}$ & $18.47^{+0.08}_{-0.07}$ & $18.58^{+0.05}_{-0.05}$ & $18.63^{+0.05}_{-0.05}$ & $10.14^{+0.08}_{-0.07}$ & $10.36^{+0.06}_{-0.06}$ & $10.51^{+0.10}_{-0.08}$ & $10.55^{+0.15}_{-0.09}$ & $2.21^{+0.09}_{-0.08}$ & $2.45^{+0.13}_{-0.10}$ & $2.61^{+0.21}_{-0.14}$ & $2.65^{+0.26}_{-0.16}$ & $22.67^{+0.12}_{-0.11}$ & $22.77^{+0.10}_{-0.09}$ & $22.84^{+0.08}_{-0.07}$ & $22.88^{+0.06}_{-0.05}$\\
Segue 1 & $18.73^{+0.53}_{-0.68}$ & $18.87^{+0.48}_{-0.67}$ & $19.00^{+0.48}_{-0.68}$ & $19.12^{+0.63}_{-0.68}$ & $9.89^{+0.72}_{-0.99}$ & $10.08^{+0.78}_{-1.01}$ & $10.21^{+0.96}_{-1.02}$ & $10.32^{+1.40}_{-1.08}$ & $1.29^{+1.15}_{-1.36}$ & $1.45^{+1.36}_{-1.38}$ & $1.57^{+1.68}_{-1.44}$ & $1.67^{+2.30}_{-1.50}$ & $23.40^{+0.47}_{-0.58}$ & $23.50^{+0.43}_{-0.54}$ & $23.59^{+0.40}_{-0.53}$ & $23.71^{+0.39}_{-0.53}$\\
Sextans I & $17.40^{+0.22}_{-0.22}$ & $17.58^{+0.18}_{-0.17}$ & $17.75^{+0.12}_{-0.11}$ & $17.87^{+0.12}_{-0.10}$ & $8.93^{+0.20}_{-0.18}$ & $9.21^{+0.15}_{-0.14}$ & $9.45^{+0.16}_{-0.13}$ & $9.56^{+0.33}_{-0.17}$ & $0.78^{+0.19}_{-0.18}$ & $1.08^{+0.21}_{-0.18}$ & $1.32^{+0.34}_{-0.22}$ & $1.42^{+0.61}_{-0.26}$ & $21.92^{+0.21}_{-0.24}$ & $22.03^{+0.19}_{-0.20}$ & $22.13^{+0.15}_{-0.15}$ & $22.21^{+0.10}_{-0.09}$\\
Tucana II & $18.56^{+0.57}_{-0.53}$ & $18.74^{+0.55}_{-0.51}$ & $18.93^{+0.56}_{-0.50}$ & $19.13^{+0.65}_{-0.56}$ & $10.40^{+0.82}_{-0.74}$ & $10.68^{+0.85}_{-0.76}$ & $10.96^{+0.92}_{-0.82}$ & $11.24^{+1.13}_{-0.99}$ & $2.57^{+1.18}_{-1.04}$ & $2.91^{+1.24}_{-1.11}$ & $3.23^{+1.36}_{-1.23}$ & $3.53^{+1.66}_{-1.44}$ & $22.91^{+0.47}_{-0.44}$ & $23.03^{+0.45}_{-0.42}$ & $23.16^{+0.43}_{-0.41}$ & $23.30^{+0.44}_{-0.39}$\\
Ursa Major I & $18.07^{+0.31}_{-0.29}$ & $18.21^{+0.28}_{-0.28}$ & $18.33^{+0.28}_{-0.28}$ & $18.40^{+0.37}_{-0.32}$ & $9.85^{+0.41}_{-0.41}$ & $10.05^{+0.45}_{-0.45}$ & $10.20^{+0.57}_{-0.51}$ & $10.26^{+0.82}_{-0.55}$ & $1.88^{+0.65}_{-0.63}$ & $2.09^{+0.78}_{-0.69}$ & $2.22^{+1.02}_{-0.73}$ & $2.27^{+1.36}_{-0.76}$ & $22.44^{+0.28}_{-0.28}$ & $22.53^{+0.25}_{-0.25}$ & $22.61^{+0.22}_{-0.22}$ & $22.68^{+0.22}_{-0.23}$\\
Ursa Major II & $19.00^{+0.45}_{-0.40}$ & $19.21^{+0.42}_{-0.39}$ & $19.44^{+0.41}_{-0.39}$ & $19.72^{+0.54}_{-0.49}$ & $10.84^{+0.61}_{-0.58}$ & $11.17^{+0.63}_{-0.61}$ & $11.51^{+0.71}_{-0.69}$ & $11.90^{+1.08}_{-0.93}$ & $3.07^{+0.93}_{-0.87}$ & $3.44^{+1.01}_{-0.94}$ & $3.83^{+1.15}_{-1.10}$ & $4.25^{+1.64}_{-1.39}$ & $23.37^{+0.39}_{-0.33}$ & $23.51^{+0.36}_{-0.32}$ & $23.66^{+0.33}_{-0.30}$ & $23.85^{+0.33}_{-0.32}$\\
Ursa Minor I & $18.56^{+0.17}_{-0.21}$ & $18.68^{+0.14}_{-0.16}$ & $18.76^{+0.12}_{-0.11}$ & $18.80^{+0.11}_{-0.11}$ & $10.35^{+0.17}_{-0.17}$ & $10.53^{+0.14}_{-0.14}$ & $10.65^{+0.15}_{-0.14}$ & $10.68^{+0.18}_{-0.14}$ & $2.39^{+0.18}_{-0.17}$ & $2.59^{+0.19}_{-0.17}$ & $2.71^{+0.24}_{-0.19}$ & $2.73^{+0.29}_{-0.19}$ & $22.92^{+0.16}_{-0.21}$ & $22.99^{+0.15}_{-0.18}$ & $23.05^{+0.13}_{-0.15}$ & $23.07^{+0.12}_{-0.12}$\\
Willman 1 & $19.11^{+0.44}_{-0.45}$ & $19.24^{+0.45}_{-0.43}$ & $19.36^{+0.52}_{-0.46}$ & $19.46^{+0.73}_{-0.52}$ & $10.73^{+0.74}_{-0.64}$ & $10.90^{+0.87}_{-0.72}$ & $11.05^{+1.11}_{-0.82}$ & $11.14^{+1.60}_{-0.89}$ & $2.58^{+1.33}_{-1.00}$ & $2.78^{+1.54}_{-1.12}$ & $2.94^{+1.86}_{-1.25}$ & $3.01^{+2.47}_{-1.32}$ & $23.54^{+0.43}_{-0.41}$ & $23.63^{+0.39}_{-0.40}$ & $23.72^{+0.38}_{-0.38}$ & $23.82^{+0.42}_{-0.39}$\\
\hline \end{tabular}

%% file: dSph_Jfactors.bbl
\begin{thebibliography}{66}%
\makeatletter
\providecommand \@ifxundefined [1]{%
 \@ifx{#1\undefined}
}%
\providecommand \@ifnum [1]{%
 \ifnum #1\expandafter \@firstoftwo
 \else \expandafter \@secondoftwo
 \fi
}%
\providecommand \@ifx [1]{%
 \ifx #1\expandafter \@firstoftwo
 \else \expandafter \@secondoftwo
 \fi
}%
\providecommand \natexlab [1]{#1}%
\providecommand \enquote  [1]{``#1''}%
\providecommand \bibnamefont  [1]{#1}%
\providecommand \bibfnamefont [1]{#1}%
\providecommand \citenamefont [1]{#1}%
\providecommand \href@noop [0]{\@secondoftwo}%
\providecommand \href [0]{\begingroup \@sanitize@url \@href}%
\providecommand \@href[1]{\@@startlink{#1}\@@href}%
\providecommand \@@href[1]{\endgroup#1\@@endlink}%
\providecommand \@sanitize@url [0]{\catcode `\\12\catcode `\$12\catcode
  `\&12\catcode `\#12\catcode `\^12\catcode `\_12\catcode `\%12\relax}%
\providecommand \@@startlink[1]{}%
\providecommand \@@endlink[0]{}%
\providecommand \url  [0]{\begingroup\@sanitize@url \@url }%
\providecommand \@url [1]{\endgroup\@href {#1}{\urlprefix }}%
\providecommand \urlprefix  [0]{URL }%
\providecommand \Eprint [0]{\href }%
\providecommand \doibase [0]{http://dx.doi.org/}%
\providecommand \selectlanguage [0]{\@gobble}%
\providecommand \bibinfo  [0]{\@secondoftwo}%
\providecommand \bibfield  [0]{\@secondoftwo}%
\providecommand \translation [1]{[#1]}%
\providecommand \BibitemOpen [0]{}%
\providecommand \bibitemStop [0]{}%
\providecommand \bibitemNoStop [0]{.\EOS\space}%
\providecommand \EOS [0]{\spacefactor3000\relax}%
\providecommand \BibitemShut  [1]{\csname bibitem#1\endcsname}%
\let\auto@bib@innerbib\@empty
\bibitem [{\citenamefont {Robertson}\ and\ \citenamefont
  {Zentner}(2009)}]{Robertson:2009bh}%
  \BibitemOpen
  \bibfield  {author} {\bibinfo {author} {\bibfnamefont {Brant}\ \bibnamefont
  {Robertson}}\ and\ \bibinfo {author} {\bibfnamefont {Andrew}\ \bibnamefont
  {Zentner}},\ }\bibfield  {title} {\enquote {\bibinfo {title} {Dark {{Matter
  Annihilation Rates}} with {{Velocity}}-{{Dependent Annihilation Cross
  Sections}}},}\ }\href {\doibase 10.1103/PhysRevD.79.083525} {\bibfield
  {journal} {\bibinfo  {journal} {Phys. Rev. D}\ }\textbf {\bibinfo {volume}
  {79}},\ \bibinfo {pages} {083525} (\bibinfo {year} {2009})},\ \Eprint
  {http://arxiv.org/abs/0902.0362} {arXiv:0902.0362} \BibitemShut {NoStop}%
\bibitem [{\citenamefont {Ferrer}\ and\ \citenamefont
  {Hunter}(2013)}]{Ferrer:2013cla}%
  \BibitemOpen
  \bibfield  {author} {\bibinfo {author} {\bibfnamefont {Francesc}\
  \bibnamefont {Ferrer}}\ and\ \bibinfo {author} {\bibfnamefont {Daniel~R.}\
  \bibnamefont {Hunter}},\ }\bibfield  {title} {\enquote {\bibinfo {title} {The
  impact of the phase-space density on the indirect detection of dark
  matter},}\ }\href {\doibase 10.1088/1475-7516/2013/09/005} {\bibfield
  {journal} {\bibinfo  {journal} {J. Cosmol. Astropart. Phys.}\ }\textbf
  {\bibinfo {volume} {2013}},\ \bibinfo {pages} {005--005} (\bibinfo {year}
  {2013})},\ \Eprint {http://arxiv.org/abs/1306.6586} {arXiv:1306.6586}
  \BibitemShut {NoStop}%
\bibitem [{\citenamefont {Boddy}\ \emph {et~al.}(2017)\citenamefont {Boddy},
  \citenamefont {Kumar}, \citenamefont {Strigari},\ and\ \citenamefont
  {Wang}}]{Boddy:2017vpe}%
  \BibitemOpen
  \bibfield  {author} {\bibinfo {author} {\bibfnamefont {Kimberly~K.}\
  \bibnamefont {Boddy}}, \bibinfo {author} {\bibfnamefont {Jason}\ \bibnamefont
  {Kumar}}, \bibinfo {author} {\bibfnamefont {Louis~E.}\ \bibnamefont
  {Strigari}}, \ and\ \bibinfo {author} {\bibfnamefont {Mei-Yu}\ \bibnamefont
  {Wang}},\ }\bibfield  {title} {\enquote {\bibinfo {title}
  {Sommerfeld-enhanced ${J}$-factors for dwarf spheroidal galaxies},}\ }\href
  {\doibase 10.1103/PhysRevD.95.123008} {\bibfield  {journal} {\bibinfo
  {journal} {Phys. Rev. D}\ }\textbf {\bibinfo {volume} {95}},\ \bibinfo
  {pages} {123008} (\bibinfo {year} {2017})},\ \Eprint
  {http://arxiv.org/abs/1702.00408} {arXiv:1702.00408} \BibitemShut {NoStop}%
\bibitem [{\citenamefont {{Zhao}}\ \emph {et~al.}(2018)\citenamefont {{Zhao}},
  \citenamefont {{Bi}}, \citenamefont {{Yin}},\ and\ \citenamefont
  {{Zhang}}}]{Zhao2017}%
  \BibitemOpen
  \bibfield  {author} {\bibinfo {author} {\bibfnamefont {Yi}~\bibnamefont
  {{Zhao}}}, \bibinfo {author} {\bibfnamefont {Xiao-Jun}\ \bibnamefont {{Bi}}},
  \bibinfo {author} {\bibfnamefont {Peng-Fei}\ \bibnamefont {{Yin}}}, \ and\
  \bibinfo {author} {\bibfnamefont {Xinmin}\ \bibnamefont {{Zhang}}},\
  }\bibfield  {title} {\enquote {\bibinfo {title} {{Constraint on the velocity
  dependent dark matter annihilation cross section from gamma-ray and kinematic
  observations of ultrafaint dwarf galaxies}},}\ }\href {\doibase
  10.1103/PhysRevD.97.063013} {\bibfield  {journal} {\bibinfo  {journal} {PRD}\
  }\textbf {\bibinfo {volume} {97}},\ \bibinfo {eid} {063013} (\bibinfo {year}
  {2018})},\ \Eprint {http://arxiv.org/abs/1711.04696} {arXiv:1711.04696
  [astro-ph.HE]} \BibitemShut {NoStop}%
\bibitem [{\citenamefont {Petac}\ \emph {et~al.}(2018)\citenamefont {Petac},
  \citenamefont {Ullio},\ and\ \citenamefont {Valli}}]{Petac:2018gue}%
  \BibitemOpen
  \bibfield  {author} {\bibinfo {author} {\bibfnamefont {Mihael}\ \bibnamefont
  {Petac}}, \bibinfo {author} {\bibfnamefont {Piero}\ \bibnamefont {Ullio}}, \
  and\ \bibinfo {author} {\bibfnamefont {Mauro}\ \bibnamefont {Valli}},\
  }\bibfield  {title} {\enquote {\bibinfo {title} {On velocity-dependent dark
  matter annihilations in dwarf satellites},}\ }\href {\doibase
  10.1088/1475-7516/2018/12/039} {\bibfield  {journal} {\bibinfo  {journal} {J.
  Cosmol. Astropart. Phys.}\ }\textbf {\bibinfo {volume} {2018}},\ \bibinfo
  {pages} {039--039} (\bibinfo {year} {2018})},\ \Eprint
  {http://arxiv.org/abs/1804.05052} {arXiv:1804.05052} \BibitemShut {NoStop}%
\bibitem [{\citenamefont {Boddy}\ \emph
  {et~al.}(2018{\natexlab{a}})\citenamefont {Boddy}, \citenamefont {Kumar},\
  and\ \citenamefont {Strigari}}]{Boddy:2018ike}%
  \BibitemOpen
  \bibfield  {author} {\bibinfo {author} {\bibfnamefont {Kimberly~K.}\
  \bibnamefont {Boddy}}, \bibinfo {author} {\bibfnamefont {Jason}\ \bibnamefont
  {Kumar}}, \ and\ \bibinfo {author} {\bibfnamefont {Louis~E.}\ \bibnamefont
  {Strigari}},\ }\bibfield  {title} {\enquote {\bibinfo {title} {The
  {{Effective J}}-{{Factor}} of the {{Galactic Center}} for
  {{Velocity}}-{{Dependent Dark Matter Annihilation}}},}\ }\href {\doibase
  10.1103/PhysRevD.98.063012} {\bibfield  {journal} {\bibinfo  {journal} {Phys.
  Rev. D}\ }\textbf {\bibinfo {volume} {98}},\ \bibinfo {pages} {063012}
  (\bibinfo {year} {2018}{\natexlab{a}})},\ \Eprint
  {http://arxiv.org/abs/1805.08379} {arXiv:1805.08379} \BibitemShut {NoStop}%
\bibitem [{\citenamefont {Lacroix}\ \emph {et~al.}(2018)\citenamefont
  {Lacroix}, \citenamefont {Stref},\ and\ \citenamefont
  {Lavalle}}]{Lacroix:2018qqh}%
  \BibitemOpen
  \bibfield  {author} {\bibinfo {author} {\bibfnamefont {Thomas}\ \bibnamefont
  {Lacroix}}, \bibinfo {author} {\bibfnamefont {Martin}\ \bibnamefont {Stref}},
  \ and\ \bibinfo {author} {\bibfnamefont {Julien}\ \bibnamefont {Lavalle}},\
  }\bibfield  {title} {\enquote {\bibinfo {title} {Anatomy of
  {{Eddington}}-like inversion methods in the context of dark matter
  searches},}\ }\href {\doibase 10.1088/1475-7516/2018/09/040} {\bibfield
  {journal} {\bibinfo  {journal} {J. Cosmol. Astropart. Phys.}\ }\textbf
  {\bibinfo {volume} {2018}},\ \bibinfo {pages} {040--040} (\bibinfo {year}
  {2018})},\ \Eprint {http://arxiv.org/abs/1805.02403} {arXiv:1805.02403}
  \BibitemShut {NoStop}%
\bibitem [{\citenamefont {Boddy}\ \emph {et~al.}(2019)\citenamefont {Boddy},
  \citenamefont {Kumar}, \citenamefont {Runburg},\ and\ \citenamefont
  {Strigari}}]{Boddy:2019wfg}%
  \BibitemOpen
  \bibfield  {author} {\bibinfo {author} {\bibfnamefont {Kimberly~K.}\
  \bibnamefont {Boddy}}, \bibinfo {author} {\bibfnamefont {Jason}\ \bibnamefont
  {Kumar}}, \bibinfo {author} {\bibfnamefont {Jack}\ \bibnamefont {Runburg}}, \
  and\ \bibinfo {author} {\bibfnamefont {Louis~E.}\ \bibnamefont {Strigari}},\
  }\bibfield  {title} {\enquote {\bibinfo {title} {{Angular distribution of
  gamma-ray emission from velocity-dependent dark matter annihilation in
  subhalos}},}\ }\href {\doibase 10.1103/PhysRevD.100.063019} {\bibfield
  {journal} {\bibinfo  {journal} {Phys. Rev. D}\ }\textbf {\bibinfo {volume}
  {100}},\ \bibinfo {pages} {063019} (\bibinfo {year} {2019})},\ \Eprint
  {http://arxiv.org/abs/1905.03431} {arXiv:1905.03431} \BibitemShut {NoStop}%
\bibitem [{\citenamefont {{Zhao}}\ \emph {et~al.}(2016)\citenamefont {{Zhao}},
  \citenamefont {{Bi}}, \citenamefont {{Jia}}, \citenamefont {{Yin}},\ and\
  \citenamefont {{Zhu}}}]{Zhao2016}%
  \BibitemOpen
  \bibfield  {author} {\bibinfo {author} {\bibfnamefont {Yi}~\bibnamefont
  {{Zhao}}}, \bibinfo {author} {\bibfnamefont {Xiao-Jun}\ \bibnamefont {{Bi}}},
  \bibinfo {author} {\bibfnamefont {Huan-Yu}\ \bibnamefont {{Jia}}}, \bibinfo
  {author} {\bibfnamefont {Peng-Fei}\ \bibnamefont {{Yin}}}, \ and\ \bibinfo
  {author} {\bibfnamefont {Feng-Rong}\ \bibnamefont {{Zhu}}},\ }\bibfield
  {title} {\enquote {\bibinfo {title} {{Constraint on the velocity dependent
  dark matter annihilation cross section from Fermi-LAT observations of dwarf
  galaxies}},}\ }\href {\doibase 10.1103/PhysRevD.93.083513} {\bibfield
  {journal} {\bibinfo  {journal} {PRD}\ }\textbf {\bibinfo {volume} {93}},\
  \bibinfo {eid} {083513} (\bibinfo {year} {2016})},\ \Eprint
  {http://arxiv.org/abs/1601.02181} {arXiv:1601.02181 [astro-ph.HE]}
  \BibitemShut {NoStop}%
\bibitem [{\citenamefont {{Bergstr{\"o}m}}\ \emph {et~al.}(2018)\citenamefont
  {{Bergstr{\"o}m}}, \citenamefont {{Catena}}, \citenamefont {{Chiappo}},
  \citenamefont {{Conrad}}, \citenamefont {{Eurenius}}, \citenamefont
  {{Eriksson}}, \citenamefont {{H{\"o}gberg}}, \citenamefont {{Larsson}},
  \citenamefont {{Olsson}}, \citenamefont {{Unger}},\ and\ \citenamefont
  {{Wadman}}}]{Bergstrom2018}%
  \BibitemOpen
  \bibfield  {author} {\bibinfo {author} {\bibfnamefont {Sebastian}\
  \bibnamefont {{Bergstr{\"o}m}}}, \bibinfo {author} {\bibfnamefont {Riccardo}\
  \bibnamefont {{Catena}}}, \bibinfo {author} {\bibfnamefont {Andrea}\
  \bibnamefont {{Chiappo}}}, \bibinfo {author} {\bibfnamefont {Jan}\
  \bibnamefont {{Conrad}}}, \bibinfo {author} {\bibfnamefont {Bj{\"o}rn}\
  \bibnamefont {{Eurenius}}}, \bibinfo {author} {\bibfnamefont {Magdalena}\
  \bibnamefont {{Eriksson}}}, \bibinfo {author} {\bibfnamefont {Michael}\
  \bibnamefont {{H{\"o}gberg}}}, \bibinfo {author} {\bibfnamefont {Susanna}\
  \bibnamefont {{Larsson}}}, \bibinfo {author} {\bibfnamefont {Emelie}\
  \bibnamefont {{Olsson}}}, \bibinfo {author} {\bibfnamefont {Andreas}\
  \bibnamefont {{Unger}}}, \ and\ \bibinfo {author} {\bibfnamefont {Rikard}\
  \bibnamefont {{Wadman}}},\ }\bibfield  {title} {\enquote {\bibinfo {title}
  {{J -factors for self-interacting dark matter in 20 dwarf spheroidal
  galaxies}},}\ }\href {\doibase 10.1103/PhysRevD.98.043017} {\bibfield
  {journal} {\bibinfo  {journal} {PRD}\ }\textbf {\bibinfo {volume} {98}},\
  \bibinfo {eid} {043017} (\bibinfo {year} {2018})},\ \Eprint
  {http://arxiv.org/abs/1712.03188} {arXiv:1712.03188 [astro-ph.CO]}
  \BibitemShut {NoStop}%
\bibitem [{\citenamefont {Widrow}(2000)}]{Widrow:2000fv}%
  \BibitemOpen
  \bibfield  {author} {\bibinfo {author} {\bibfnamefont {Lawrence~M.}\
  \bibnamefont {Widrow}},\ }\bibfield  {title} {\enquote {\bibinfo {title}
  {{Distribution Functions for Cuspy Dark Matter Density Profiles}},}\ }\href
  {\doibase 10.1086/317367} {\bibfield  {journal} {\bibinfo  {journal}
  {Astrophys. J. Suppl.}\ }\textbf {\bibinfo {volume} {131}},\ \bibinfo {pages}
  {39--46} (\bibinfo {year} {2000})}\BibitemShut {NoStop}%
\bibitem [{\citenamefont {{Boddy}}\ \emph {et~al.}(2019)\citenamefont
  {{Boddy}}, \citenamefont {{Hill}}, \citenamefont {{Kumar}}, \citenamefont
  {{Sand ick}},\ and\ \citenamefont {{Haghi}}}]{MADHAT}%
  \BibitemOpen
  \bibfield  {author} {\bibinfo {author} {\bibfnamefont {Kimberly~K.}\
  \bibnamefont {{Boddy}}}, \bibinfo {author} {\bibfnamefont {Stephen}\
  \bibnamefont {{Hill}}}, \bibinfo {author} {\bibfnamefont {Jason}\
  \bibnamefont {{Kumar}}}, \bibinfo {author} {\bibfnamefont {Pearl}\
  \bibnamefont {{Sand ick}}}, \ and\ \bibinfo {author} {\bibfnamefont {Barmak
  Shams~Es}\ \bibnamefont {{Haghi}}},\ }\bibfield  {title} {\enquote {\bibinfo
  {title} {{MADHAT: Model-Agnostic Dark Halo Analysis Tool}},}\ }\href@noop {}
  {\ ,\ \bibinfo {eid} {arXiv:1910.02890} (\bibinfo {year} {2019})},\ \Eprint
  {http://arxiv.org/abs/1910.02890} {arXiv:1910.02890 [hep-ph]} \BibitemShut
  {NoStop}%
\bibitem [{\citenamefont {Atwood}\ \emph {et~al.}(2009)\citenamefont {Atwood}
  \emph {et~al.}}]{Atwood:2009ez}%
  \BibitemOpen
  \bibfield  {author} {\bibinfo {author} {\bibfnamefont {W.~B.}\ \bibnamefont
  {Atwood}} \emph {et~al.} (\bibinfo {collaboration} {Fermi-LAT}),\ }\bibfield
  {title} {\enquote {\bibinfo {title} {{The Large Area Telescope on the Fermi
  Gamma-ray Space Telescope Mission}},}\ }\href {\doibase
  10.1088/0004-637X/697/2/1071} {\bibfield  {journal} {\bibinfo  {journal}
  {Astrophys. J.}\ }\textbf {\bibinfo {volume} {697}},\ \bibinfo {pages}
  {1071--1102} (\bibinfo {year} {2009})},\ \Eprint
  {http://arxiv.org/abs/0902.1089} {arXiv:0902.1089 [astro-ph.IM]} \BibitemShut
  {NoStop}%
\bibitem [{\citenamefont {Boddy}\ \emph
  {et~al.}(2018{\natexlab{b}})\citenamefont {Boddy}, \citenamefont {Kumar},
  \citenamefont {Marfatia},\ and\ \citenamefont {Sandick}}]{Boddy:2018qur}%
  \BibitemOpen
  \bibfield  {author} {\bibinfo {author} {\bibfnamefont {Kimberly~K.}\
  \bibnamefont {Boddy}}, \bibinfo {author} {\bibfnamefont {Jason}\ \bibnamefont
  {Kumar}}, \bibinfo {author} {\bibfnamefont {Danny}\ \bibnamefont {Marfatia}},
  \ and\ \bibinfo {author} {\bibfnamefont {Pearl}\ \bibnamefont {Sandick}},\
  }\bibfield  {title} {\enquote {\bibinfo {title} {Model-independent
  constraints on dark matter annihilation in dwarf spheroidal galaxies},}\
  }\href {\doibase 10.1103/PhysRevD.97.095031} {\bibfield  {journal} {\bibinfo
  {journal} {Phys. Rev. D}\ }\textbf {\bibinfo {volume} {97}},\ \bibinfo
  {pages} {095031} (\bibinfo {year} {2018}{\natexlab{b}})},\ \Eprint
  {http://arxiv.org/abs/1802.03826} {arXiv:1802.03826} \BibitemShut {NoStop}%
\bibitem [{\citenamefont {{Arkani-Hamed}}\ \emph {et~al.}(2009)\citenamefont
  {{Arkani-Hamed}}, \citenamefont {Finkbeiner}, \citenamefont {Slatyer},\ and\
  \citenamefont {Weiner}}]{ArkaniHamed:2008qn}%
  \BibitemOpen
  \bibfield  {author} {\bibinfo {author} {\bibfnamefont {Nima}\ \bibnamefont
  {{Arkani-Hamed}}}, \bibinfo {author} {\bibfnamefont {Douglas~P.}\
  \bibnamefont {Finkbeiner}}, \bibinfo {author} {\bibfnamefont {Tracy~R.}\
  \bibnamefont {Slatyer}}, \ and\ \bibinfo {author} {\bibfnamefont {Neal}\
  \bibnamefont {Weiner}},\ }\bibfield  {title} {\enquote {\bibinfo {title} {A
  {{Theory}} of {{Dark Matter}}},}\ }\href {\doibase
  10.1103/PhysRevD.79.015014} {\bibfield  {journal} {\bibinfo  {journal} {Phys.
  Rev. D}\ }\textbf {\bibinfo {volume} {79}},\ \bibinfo {pages} {015014}
  (\bibinfo {year} {2009})},\ \Eprint {http://arxiv.org/abs/0810.0713}
  {arXiv:0810.0713} \BibitemShut {NoStop}%
\bibitem [{\citenamefont {Feng}\ \emph {et~al.}(2010)\citenamefont {Feng},
  \citenamefont {Kaplinghat},\ and\ \citenamefont {Yu}}]{Feng:2010zp}%
  \BibitemOpen
  \bibfield  {author} {\bibinfo {author} {\bibfnamefont {Jonathan~L.}\
  \bibnamefont {Feng}}, \bibinfo {author} {\bibfnamefont {Manoj}\ \bibnamefont
  {Kaplinghat}}, \ and\ \bibinfo {author} {\bibfnamefont {Hai-Bo}\ \bibnamefont
  {Yu}},\ }\bibfield  {title} {\enquote {\bibinfo {title} {Sommerfeld
  {{Enhancements}} for {{Thermal Relic Dark Matter}}},}\ }\href {\doibase
  10.1103/PhysRevD.82.083525} {\bibfield  {journal} {\bibinfo  {journal} {Phys.
  Rev. D}\ }\textbf {\bibinfo {volume} {82}},\ \bibinfo {pages} {083525}
  (\bibinfo {year} {2010})},\ \Eprint {http://arxiv.org/abs/1005.4678}
  {arXiv:1005.4678} \BibitemShut {NoStop}%
\bibitem [{\citenamefont {Kumar}\ and\ \citenamefont
  {Marfatia}(2013)}]{Kumar:2013iva}%
  \BibitemOpen
  \bibfield  {author} {\bibinfo {author} {\bibfnamefont {Jason}\ \bibnamefont
  {Kumar}}\ and\ \bibinfo {author} {\bibfnamefont {Danny}\ \bibnamefont
  {Marfatia}},\ }\bibfield  {title} {\enquote {\bibinfo {title} {Matrix element
  analyses of dark matter scattering and annihilation},}\ }\href {\doibase
  10.1103/PhysRevD.88.014035} {\bibfield  {journal} {\bibinfo  {journal} {Phys.
  Rev. D}\ }\textbf {\bibinfo {volume} {88}},\ \bibinfo {pages} {014035}
  (\bibinfo {year} {2013})},\ \Eprint {http://arxiv.org/abs/1305.1611}
  {arXiv:1305.1611} \BibitemShut {NoStop}%
\bibitem [{\citenamefont {Giacchino}\ \emph {et~al.}(2013)\citenamefont
  {Giacchino}, \citenamefont {{Lopez-Honorez}},\ and\ \citenamefont
  {Tytgat}}]{Giacchino:2013bta}%
  \BibitemOpen
  \bibfield  {author} {\bibinfo {author} {\bibfnamefont {Federica}\
  \bibnamefont {Giacchino}}, \bibinfo {author} {\bibfnamefont {Laura}\
  \bibnamefont {{Lopez-Honorez}}}, \ and\ \bibinfo {author} {\bibfnamefont
  {Michel H.~G.}\ \bibnamefont {Tytgat}},\ }\bibfield  {title} {\enquote
  {\bibinfo {title} {Scalar {{Dark Matter Models}} with {{Significant Internal
  Bremsstrahlung}}},}\ }\href {\doibase 10.1088/1475-7516/2013/10/025}
  {\bibfield  {journal} {\bibinfo  {journal} {J. Cosmol. Astropart. Phys.}\
  }\textbf {\bibinfo {volume} {2013}},\ \bibinfo {pages} {025--025} (\bibinfo
  {year} {2013})},\ \Eprint {http://arxiv.org/abs/1307.6480} {arXiv:1307.6480}
  \BibitemShut {NoStop}%
\bibitem [{\citenamefont {{Pace}}\ and\ \citenamefont
  {{Strigari}}(2019)}]{Pace:2018tin}%
  \BibitemOpen
  \bibfield  {author} {\bibinfo {author} {\bibfnamefont {Andrew~B.}\
  \bibnamefont {{Pace}}}\ and\ \bibinfo {author} {\bibfnamefont {Louis~E.}\
  \bibnamefont {{Strigari}}},\ }\bibfield  {title} {\enquote {\bibinfo {title}
  {{Scaling relations for dark matter annihilation and decay profiles in dwarf
  spheroidal galaxies}},}\ }\href {\doibase 10.1093/mnras/sty2839} {\bibfield
  {journal} {\bibinfo  {journal} {MNRAS}\ }\textbf {\bibinfo {volume} {482}},\
  \bibinfo {pages} {3480--3496} (\bibinfo {year} {2019})},\ \Eprint
  {http://arxiv.org/abs/1802.06811} {arXiv:1802.06811 [astro-ph.GA]}
  \BibitemShut {NoStop}%
\bibitem [{\citenamefont {{Strigari}}\ \emph {et~al.}(2008)\citenamefont
  {{Strigari}}, \citenamefont {{Koushiappas}}, \citenamefont {{Bullock}},
  \citenamefont {{Kaplinghat}}, \citenamefont {{Simon}}, \citenamefont
  {{Geha}},\ and\ \citenamefont {{Willman}}}]{Strigari2008ApJ...678..614S}%
  \BibitemOpen
  \bibfield  {author} {\bibinfo {author} {\bibfnamefont {Louis~E.}\
  \bibnamefont {{Strigari}}}, \bibinfo {author} {\bibfnamefont {Savvas~M.}\
  \bibnamefont {{Koushiappas}}}, \bibinfo {author} {\bibfnamefont {James~S.}\
  \bibnamefont {{Bullock}}}, \bibinfo {author} {\bibfnamefont {Manoj}\
  \bibnamefont {{Kaplinghat}}}, \bibinfo {author} {\bibfnamefont {Joshua~D.}\
  \bibnamefont {{Simon}}}, \bibinfo {author} {\bibfnamefont {Marla}\
  \bibnamefont {{Geha}}}, \ and\ \bibinfo {author} {\bibfnamefont {Beth}\
  \bibnamefont {{Willman}}},\ }\bibfield  {title} {\enquote {\bibinfo {title}
  {{The Most Dark-Matter-dominated Galaxies: Predicted Gamma-Ray Signals from
  the Faintest Milky Way Dwarfs}},}\ }\href {\doibase 10.1086/529488}
  {\bibfield  {journal} {\bibinfo  {journal} {ApJ}\ }\textbf {\bibinfo {volume}
  {678}},\ \bibinfo {pages} {614--620} (\bibinfo {year} {2008})},\ \Eprint
  {http://arxiv.org/abs/0709.1510} {arXiv:0709.1510 [astro-ph]} \BibitemShut
  {NoStop}%
\bibitem [{\citenamefont {{Bonnivard}}\ \emph
  {et~al.}(2015{\natexlab{a}})\citenamefont {{Bonnivard}}, \citenamefont
  {{Combet}}, \citenamefont {{Daniel}}, \citenamefont {{Funk}}, \citenamefont
  {{Geringer-Sameth}}, \citenamefont {{Hinton}}, \citenamefont {{Maurin}},
  \citenamefont {{Read}}, \citenamefont {{Sarkar}}, \citenamefont {{Walker}},\
  and\ \citenamefont {{Wilkinson}}}]{Bonnivard2015MNRAS.453..849B}%
  \BibitemOpen
  \bibfield  {author} {\bibinfo {author} {\bibfnamefont {V.}~\bibnamefont
  {{Bonnivard}}}, \bibinfo {author} {\bibfnamefont {C.}~\bibnamefont
  {{Combet}}}, \bibinfo {author} {\bibfnamefont {M.}~\bibnamefont {{Daniel}}},
  \bibinfo {author} {\bibfnamefont {S.}~\bibnamefont {{Funk}}}, \bibinfo
  {author} {\bibfnamefont {A.}~\bibnamefont {{Geringer-Sameth}}}, \bibinfo
  {author} {\bibfnamefont {J.~A.}\ \bibnamefont {{Hinton}}}, \bibinfo {author}
  {\bibfnamefont {D.}~\bibnamefont {{Maurin}}}, \bibinfo {author}
  {\bibfnamefont {J.~I.}\ \bibnamefont {{Read}}}, \bibinfo {author}
  {\bibfnamefont {S.}~\bibnamefont {{Sarkar}}}, \bibinfo {author}
  {\bibfnamefont {M.~G.}\ \bibnamefont {{Walker}}}, \ and\ \bibinfo {author}
  {\bibfnamefont {M.~I.}\ \bibnamefont {{Wilkinson}}},\ }\bibfield  {title}
  {\enquote {\bibinfo {title} {{Dark matter annihilation and decay in dwarf
  spheroidal galaxies: the classical and ultrafaint dSphs}},}\ }\href {\doibase
  10.1093/mnras/stv1601} {\bibfield  {journal} {\bibinfo  {journal} {MNRAS}\
  }\textbf {\bibinfo {volume} {453}},\ \bibinfo {pages} {849--867} (\bibinfo
  {year} {2015}{\natexlab{a}})},\ \Eprint {http://arxiv.org/abs/1504.02048}
  {arXiv:1504.02048 [astro-ph.HE]} \BibitemShut {NoStop}%
\bibitem [{\citenamefont {{Geringer-Sameth}}\ \emph
  {et~al.}(2015{\natexlab{a}})\citenamefont {{Geringer-Sameth}}, \citenamefont
  {{Koushiappas}},\ and\ \citenamefont
  {{Walker}}}]{Geringer-Sameth2015ApJ...801...74G}%
  \BibitemOpen
  \bibfield  {author} {\bibinfo {author} {\bibfnamefont {A.}~\bibnamefont
  {{Geringer-Sameth}}}, \bibinfo {author} {\bibfnamefont {S.~M.}\ \bibnamefont
  {{Koushiappas}}}, \ and\ \bibinfo {author} {\bibfnamefont {M.}~\bibnamefont
  {{Walker}}},\ }\bibfield  {title} {\enquote {\bibinfo {title} {{Dwarf Galaxy
  Annihilation and Decay Emission Profiles for Dark Matter Experiments}},}\
  }\href {\doibase 10.1088/0004-637X/801/2/74} {\bibfield  {journal} {\bibinfo
  {journal} {ApJ}\ }\textbf {\bibinfo {volume} {801}},\ \bibinfo {eid} {74}
  (\bibinfo {year} {2015}{\natexlab{a}})},\ \Eprint
  {http://arxiv.org/abs/1408.0002} {arXiv:1408.0002} \BibitemShut {NoStop}%
\bibitem [{\citenamefont {Plummer}(1911)}]{Plummer1911MNRAS..71..460P}%
  \BibitemOpen
  \bibfield  {author} {\bibinfo {author} {\bibfnamefont {Henry~C.}\
  \bibnamefont {Plummer}},\ }\bibfield  {title} {\enquote {\bibinfo {title}
  {{On the problem of distribution in globular star clusters}},}\ }\href@noop
  {} {\bibfield  {journal} {\bibinfo  {journal} {Mon. Not. Roy. Astron. Soc.}\
  }\textbf {\bibinfo {volume} {71}},\ \bibinfo {pages} {460--470} (\bibinfo
  {year} {1911})}\BibitemShut {NoStop}%
\bibitem [{\citenamefont {{Navarro}}\ \emph {et~al.}(1997)\citenamefont
  {{Navarro}}, \citenamefont {{Frenk}},\ and\ \citenamefont
  {{White}}}]{Navarro1997ApJ...490..493N}%
  \BibitemOpen
  \bibfield  {author} {\bibinfo {author} {\bibfnamefont {J.~F.}\ \bibnamefont
  {{Navarro}}}, \bibinfo {author} {\bibfnamefont {C.~S.}\ \bibnamefont
  {{Frenk}}}, \ and\ \bibinfo {author} {\bibfnamefont {S.~D.~M.}\ \bibnamefont
  {{White}}},\ }\bibfield  {title} {\enquote {\bibinfo {title} {{A Universal
  Density Profile from Hierarchical Clustering}},}\ }\href {\doibase
  10.1086/304888} {\bibfield  {journal} {\bibinfo  {journal} {ApJ}\ }\textbf
  {\bibinfo {volume} {490}},\ \bibinfo {pages} {493--508} (\bibinfo {year}
  {1997})},\ \Eprint {http://arxiv.org/abs/astro-ph/9611107} {astro-ph/9611107}
  \BibitemShut {NoStop}%
\bibitem [{\citenamefont {{Evans}}\ \emph {et~al.}(2009)\citenamefont
  {{Evans}}, \citenamefont {{An}},\ and\ \citenamefont
  {{Walker}}}]{Evans2009MNRAS.393L..50E}%
  \BibitemOpen
  \bibfield  {author} {\bibinfo {author} {\bibfnamefont {N.~W.}\ \bibnamefont
  {{Evans}}}, \bibinfo {author} {\bibfnamefont {J.}~\bibnamefont {{An}}}, \
  and\ \bibinfo {author} {\bibfnamefont {M.~G.}\ \bibnamefont {{Walker}}},\
  }\bibfield  {title} {\enquote {\bibinfo {title} {{Cores and cusps in the
  dwarf spheroidals}},}\ }\href {\doibase 10.1111/j.1745-3933.2008.00596.x}
  {\bibfield  {journal} {\bibinfo  {journal} {Monthly Notices of the Royal
  Astronomical Society}\ }\textbf {\bibinfo {volume} {393}},\ \bibinfo {pages}
  {L50--L54} (\bibinfo {year} {2009})},\ \Eprint
  {http://arxiv.org/abs/0811.1488} {arXiv:0811.1488 [astro-ph]} \BibitemShut
  {NoStop}%
\bibitem [{\citenamefont {{An}}\ and\ \citenamefont
  {{Evans}}(2009)}]{An2009ApJ...701.1500A}%
  \BibitemOpen
  \bibfield  {author} {\bibinfo {author} {\bibfnamefont {Jin~H.}\ \bibnamefont
  {{An}}}\ and\ \bibinfo {author} {\bibfnamefont {N.~Wyn}\ \bibnamefont
  {{Evans}}},\ }\bibfield  {title} {\enquote {\bibinfo {title} {{A Theorem on
  Central Velocity Dispersions}},}\ }\href {\doibase
  10.1088/0004-637X/701/2/1500} {\bibfield  {journal} {\bibinfo  {journal}
  {ApJ}\ }\textbf {\bibinfo {volume} {701}},\ \bibinfo {pages} {1500--1505}
  (\bibinfo {year} {2009})},\ \Eprint {http://arxiv.org/abs/0906.3673}
  {arXiv:0906.3673 [astro-ph.GA]} \BibitemShut {NoStop}%
\bibitem [{\citenamefont {{Ciotti}}\ and\ \citenamefont
  {{Morganti}}(2010)}]{Ciotti2010MNRAS.408.1070C}%
  \BibitemOpen
  \bibfield  {author} {\bibinfo {author} {\bibfnamefont {Luca}\ \bibnamefont
  {{Ciotti}}}\ and\ \bibinfo {author} {\bibfnamefont {Lucia}\ \bibnamefont
  {{Morganti}}},\ }\bibfield  {title} {\enquote {\bibinfo {title} {{How general
  is the global density slope-anisotropy inequality?}}}\ }\href {\doibase
  10.1111/j.1365-2966.2010.17184.x} {\bibfield  {journal} {\bibinfo  {journal}
  {Monthly Notices of the Royal Astronomical Society}\ }\textbf {\bibinfo
  {volume} {408}},\ \bibinfo {pages} {1070--1074} (\bibinfo {year} {2010})},\
  \Eprint {http://arxiv.org/abs/1006.2344} {arXiv:1006.2344 [astro-ph.CO]}
  \BibitemShut {NoStop}%
\bibitem [{\citenamefont {{Bonnivard}}\ \emph
  {et~al.}(2015{\natexlab{b}})\citenamefont {{Bonnivard}}, \citenamefont
  {{Combet}}, \citenamefont {{Maurin}},\ and\ \citenamefont
  {{Walker}}}]{Bonnivard2015MNRAS.446.3002B}%
  \BibitemOpen
  \bibfield  {author} {\bibinfo {author} {\bibfnamefont {V.}~\bibnamefont
  {{Bonnivard}}}, \bibinfo {author} {\bibfnamefont {C.}~\bibnamefont
  {{Combet}}}, \bibinfo {author} {\bibfnamefont {D.}~\bibnamefont {{Maurin}}},
  \ and\ \bibinfo {author} {\bibfnamefont {M.~G.}\ \bibnamefont {{Walker}}},\
  }\bibfield  {title} {\enquote {\bibinfo {title} {{Spherical Jeans analysis
  for dark matter indirect detection in dwarf spheroidal galaxies - impact of
  physical parameters and triaxiality}},}\ }\href {\doibase
  10.1093/mnras/stu2296} {\bibfield  {journal} {\bibinfo  {journal} {Monthly
  Notices of the Royal Astronomical Society}\ }\textbf {\bibinfo {volume}
  {446}},\ \bibinfo {pages} {3002--3021} (\bibinfo {year}
  {2015}{\natexlab{b}})},\ \Eprint {http://arxiv.org/abs/1407.7822}
  {arXiv:1407.7822 [astro-ph.HE]} \BibitemShut {NoStop}%
\bibitem [{\citenamefont {{Feroz}}\ and\ \citenamefont
  {{Hobson}}(2008)}]{Feroz2008MNRAS.384..449F}%
  \BibitemOpen
  \bibfield  {author} {\bibinfo {author} {\bibfnamefont {F.}~\bibnamefont
  {{Feroz}}}\ and\ \bibinfo {author} {\bibfnamefont {M.~P.}\ \bibnamefont
  {{Hobson}}},\ }\bibfield  {title} {\enquote {\bibinfo {title} {{Multimodal
  nested sampling: an efficient and robust alternative to Markov Chain Monte
  Carlo methods for astronomical data analyses}},}\ }\href {\doibase
  10.1111/j.1365-2966.2007.12353.x} {\bibfield  {journal} {\bibinfo  {journal}
  {MNRAS}\ }\textbf {\bibinfo {volume} {384}},\ \bibinfo {pages} {449--463}
  (\bibinfo {year} {2008})},\ \Eprint {http://arxiv.org/abs/0704.3704}
  {arXiv:0704.3704} \BibitemShut {NoStop}%
\bibitem [{\citenamefont {{Feroz}}\ \emph {et~al.}(2009)\citenamefont
  {{Feroz}}, \citenamefont {{Hobson}},\ and\ \citenamefont
  {{Bridges}}}]{Feroz2009MNRAS.398.1601F}%
  \BibitemOpen
  \bibfield  {author} {\bibinfo {author} {\bibfnamefont {F.}~\bibnamefont
  {{Feroz}}}, \bibinfo {author} {\bibfnamefont {M.~P.}\ \bibnamefont
  {{Hobson}}}, \ and\ \bibinfo {author} {\bibfnamefont {M.}~\bibnamefont
  {{Bridges}}},\ }\bibfield  {title} {\enquote {\bibinfo {title} {{MULTINEST:
  an efficient and robust Bayesian inference tool for cosmology and particle
  physics}},}\ }\href {\doibase 10.1111/j.1365-2966.2009.14548.x} {\bibfield
  {journal} {\bibinfo  {journal} {MNRAS}\ }\textbf {\bibinfo {volume} {398}},\
  \bibinfo {pages} {1601--1614} (\bibinfo {year} {2009})},\ \Eprint
  {http://arxiv.org/abs/0809.3437} {arXiv:0809.3437} \BibitemShut {NoStop}%
\bibitem [{\citenamefont {{Caldwell}}\ \emph {et~al.}(2017)\citenamefont
  {{Caldwell}}, \citenamefont {{Walker}}, \citenamefont {{Mateo}},
  \citenamefont {{Olszewski}}, \citenamefont {{Koposov}}, \citenamefont
  {{Belokurov}}, \citenamefont {{Torrealba}}, \citenamefont
  {{Geringer-Sameth}},\ and\ \citenamefont {{Johnson}}}]{Caldwell:2016hrl}%
  \BibitemOpen
  \bibfield  {author} {\bibinfo {author} {\bibfnamefont {N.}~\bibnamefont
  {{Caldwell}}}, \bibinfo {author} {\bibfnamefont {M.~G.}\ \bibnamefont
  {{Walker}}}, \bibinfo {author} {\bibfnamefont {M.}~\bibnamefont {{Mateo}}},
  \bibinfo {author} {\bibfnamefont {E.~W.}\ \bibnamefont {{Olszewski}}},
  \bibinfo {author} {\bibfnamefont {S.}~\bibnamefont {{Koposov}}}, \bibinfo
  {author} {\bibfnamefont {V.}~\bibnamefont {{Belokurov}}}, \bibinfo {author}
  {\bibfnamefont {G.}~\bibnamefont {{Torrealba}}}, \bibinfo {author}
  {\bibfnamefont {A.}~\bibnamefont {{Geringer-Sameth}}}, \ and\ \bibinfo
  {author} {\bibfnamefont {C.~I.}\ \bibnamefont {{Johnson}}},\ }\bibfield
  {title} {\enquote {\bibinfo {title} {{Crater 2: An Extremely Cold Dark Matter
  Halo}},}\ }\href {\doibase 10.3847/1538-4357/aa688e} {\bibfield  {journal}
  {\bibinfo  {journal} {ApJ}\ }\textbf {\bibinfo {volume} {839}},\ \bibinfo
  {eid} {20} (\bibinfo {year} {2017})},\ \Eprint
  {http://arxiv.org/abs/1612.06398} {arXiv:1612.06398} \BibitemShut {NoStop}%
\bibitem [{\citenamefont {{Koposov}}\ \emph {et~al.}(2018)\citenamefont
  {{Koposov}}, \citenamefont {{Walker}}, \citenamefont {{Belokurov}},
  \citenamefont {{Casey}}, \citenamefont {{Geringer-Sameth}}, \citenamefont
  {{Mackey}}, \citenamefont {{Da Costa}}, \citenamefont {{Erkal}},
  \citenamefont {{Jethwa}}, \citenamefont {{Mateo}}, \citenamefont
  {{Olszewski}},\ and\ \citenamefont {{Bailey}}}]{Koposov2018MNRAS.479.5343K}%
  \BibitemOpen
  \bibfield  {author} {\bibinfo {author} {\bibfnamefont {S.~E.}\ \bibnamefont
  {{Koposov}}}, \bibinfo {author} {\bibfnamefont {M.~G.}\ \bibnamefont
  {{Walker}}}, \bibinfo {author} {\bibfnamefont {V.}~\bibnamefont
  {{Belokurov}}}, \bibinfo {author} {\bibfnamefont {A.~R.}\ \bibnamefont
  {{Casey}}}, \bibinfo {author} {\bibfnamefont {A.}~\bibnamefont
  {{Geringer-Sameth}}}, \bibinfo {author} {\bibfnamefont {D.}~\bibnamefont
  {{Mackey}}}, \bibinfo {author} {\bibfnamefont {G.}~\bibnamefont {{Da
  Costa}}}, \bibinfo {author} {\bibfnamefont {D.}~\bibnamefont {{Erkal}}},
  \bibinfo {author} {\bibfnamefont {P.}~\bibnamefont {{Jethwa}}}, \bibinfo
  {author} {\bibfnamefont {M.}~\bibnamefont {{Mateo}}}, \bibinfo {author}
  {\bibfnamefont {E.~W.}\ \bibnamefont {{Olszewski}}}, \ and\ \bibinfo {author}
  {\bibfnamefont {J.~I.}\ \bibnamefont {{Bailey}}},\ }\bibfield  {title}
  {\enquote {\bibinfo {title} {{Snake in the Clouds: a new nearby dwarf galaxy
  in the Magellanic bridge*}},}\ }\href {\doibase 10.1093/mnras/sty1772}
  {\bibfield  {journal} {\bibinfo  {journal} {MNRAS}\ }\textbf {\bibinfo
  {volume} {479}},\ \bibinfo {pages} {5343--5361} (\bibinfo {year} {2018})},\
  \Eprint {http://arxiv.org/abs/1804.06430} {arXiv:1804.06430} \BibitemShut
  {NoStop}%
\bibitem [{\citenamefont {{Laevens}}\ \emph {et~al.}(2015)\citenamefont
  {{Laevens}}, \citenamefont {{Martin}}, \citenamefont {{Bernard}},
  \citenamefont {{Schlafly}}, \citenamefont {{Sesar}}, \citenamefont {{Rix}},
  \citenamefont {{Bell}}, \citenamefont {{Ferguson}}, \citenamefont {{Slater}},
  \citenamefont {{Sweeney}}, \citenamefont {{Wyse}}, \citenamefont {{Huxor}},
  \citenamefont {{Burgett}}, \citenamefont {{Chambers}}, \citenamefont
  {{Draper}}, \citenamefont {{Hodapp}}, \citenamefont {{Kaiser}}, \citenamefont
  {{Magnier}}, \citenamefont {{Metcalfe}}, \citenamefont {{Tonry}},
  \citenamefont {{Wainscoat}},\ and\ \citenamefont
  {{Waters}}}]{Laevens:2015kla}%
  \BibitemOpen
  \bibfield  {author} {\bibinfo {author} {\bibfnamefont {Benjamin P.~M.}\
  \bibnamefont {{Laevens}}}, \bibinfo {author} {\bibfnamefont {Nicolas~F.}\
  \bibnamefont {{Martin}}}, \bibinfo {author} {\bibfnamefont {Edouard~J.}\
  \bibnamefont {{Bernard}}}, \bibinfo {author} {\bibfnamefont {Edward~F.}\
  \bibnamefont {{Schlafly}}}, \bibinfo {author} {\bibfnamefont {Branimir}\
  \bibnamefont {{Sesar}}}, \bibinfo {author} {\bibfnamefont {Hans-Walter}\
  \bibnamefont {{Rix}}}, \bibinfo {author} {\bibfnamefont {Eric~F.}\
  \bibnamefont {{Bell}}}, \bibinfo {author} {\bibfnamefont {Annette M.~N.}\
  \bibnamefont {{Ferguson}}}, \bibinfo {author} {\bibfnamefont {Colin~T.}\
  \bibnamefont {{Slater}}}, \bibinfo {author} {\bibfnamefont {William~E.}\
  \bibnamefont {{Sweeney}}}, \bibinfo {author} {\bibfnamefont {Rosemary F.~G.}\
  \bibnamefont {{Wyse}}}, \bibinfo {author} {\bibfnamefont {Avon~P.}\
  \bibnamefont {{Huxor}}}, \bibinfo {author} {\bibfnamefont {William~S.}\
  \bibnamefont {{Burgett}}}, \bibinfo {author} {\bibfnamefont {Kenneth~C.}\
  \bibnamefont {{Chambers}}}, \bibinfo {author} {\bibfnamefont {Peter~W.}\
  \bibnamefont {{Draper}}}, \bibinfo {author} {\bibfnamefont {Klaus~A.}\
  \bibnamefont {{Hodapp}}}, \bibinfo {author} {\bibfnamefont {Nicholas}\
  \bibnamefont {{Kaiser}}}, \bibinfo {author} {\bibfnamefont {Eugene~A.}\
  \bibnamefont {{Magnier}}}, \bibinfo {author} {\bibfnamefont {Nigel}\
  \bibnamefont {{Metcalfe}}}, \bibinfo {author} {\bibfnamefont {John~L.}\
  \bibnamefont {{Tonry}}}, \bibinfo {author} {\bibfnamefont {Richard~J.}\
  \bibnamefont {{Wainscoat}}}, \ and\ \bibinfo {author} {\bibfnamefont
  {Christopher}\ \bibnamefont {{Waters}}},\ }\bibfield  {title} {\enquote
  {\bibinfo {title} {{Sagittarius II, Draco II and Laevens 3: Three New Milky
  Way Satellites Discovered in the Pan-STARRS 1 3{\ensuremath{\pi}} Survey}},}\
  }\href {\doibase 10.1088/0004-637X/813/1/44} {\bibfield  {journal} {\bibinfo
  {journal} {ApJ}\ }\textbf {\bibinfo {volume} {813}},\ \bibinfo {eid} {44}
  (\bibinfo {year} {2015})},\ \Eprint {http://arxiv.org/abs/1507.07564}
  {arXiv:1507.07564 [astro-ph.GA]} \BibitemShut {NoStop}%
\bibitem [{\citenamefont {{Mutlu-Pakdil}}\ \emph {et~al.}(2018)\citenamefont
  {{Mutlu-Pakdil}}, \citenamefont {{Sand}}, \citenamefont {{Carlin}},
  \citenamefont {{Spekkens}}, \citenamefont {{Caldwell}}, \citenamefont
  {{Crnojevi{\'c}}}, \citenamefont {{Hughes}}, \citenamefont {{Willman}},\ and\
  \citenamefont {{Zaritsky}}}]{MutlaPakdil2018ApJ...863...25M}%
  \BibitemOpen
  \bibfield  {author} {\bibinfo {author} {\bibfnamefont {B.}~\bibnamefont
  {{Mutlu-Pakdil}}}, \bibinfo {author} {\bibfnamefont {D.~J.}\ \bibnamefont
  {{Sand}}}, \bibinfo {author} {\bibfnamefont {J.~L.}\ \bibnamefont
  {{Carlin}}}, \bibinfo {author} {\bibfnamefont {K.}~\bibnamefont
  {{Spekkens}}}, \bibinfo {author} {\bibfnamefont {N.}~\bibnamefont
  {{Caldwell}}}, \bibinfo {author} {\bibfnamefont {D.}~\bibnamefont
  {{Crnojevi{\'c}}}}, \bibinfo {author} {\bibfnamefont {A.~K.}\ \bibnamefont
  {{Hughes}}}, \bibinfo {author} {\bibfnamefont {B.}~\bibnamefont {{Willman}}},
  \ and\ \bibinfo {author} {\bibfnamefont {D.}~\bibnamefont {{Zaritsky}}},\
  }\bibfield  {title} {\enquote {\bibinfo {title} {{A Deeper Look at the New
  Milky Way Satellites: Sagittarius II, Reticulum II, Phoenix II, and Tucana
  III}},}\ }\href {\doibase 10.3847/1538-4357/aacd0e} {\bibfield  {journal}
  {\bibinfo  {journal} {ApJ}\ }\textbf {\bibinfo {volume} {863}},\ \bibinfo
  {eid} {25} (\bibinfo {year} {2018})},\ \Eprint
  {http://arxiv.org/abs/1804.08627} {arXiv:1804.08627} \BibitemShut {NoStop}%
\bibitem [{\citenamefont {{Longeard}}\ \emph {et~al.}(2019)\citenamefont
  {{Longeard}}, \citenamefont {{Martin}}, \citenamefont {{Starkenburg}},
  \citenamefont {{Ibata}}, \citenamefont {{Collins}}, \citenamefont
  {{Laevens}}, \citenamefont {{Mackey}}, \citenamefont {{Rich}}, \citenamefont
  {{Aguado}}, \citenamefont {{Arentsen}}, \citenamefont {{Jablonka}},
  \citenamefont {{Gonzalez Hernandez}}, \citenamefont {{Navarro}},\ and\
  \citenamefont {{Sanchez-Janssen}}}]{Longeard2019arXiv190202780L}%
  \BibitemOpen
  \bibfield  {author} {\bibinfo {author} {\bibfnamefont {N.}~\bibnamefont
  {{Longeard}}}, \bibinfo {author} {\bibfnamefont {N.}~\bibnamefont
  {{Martin}}}, \bibinfo {author} {\bibfnamefont {E.}~\bibnamefont
  {{Starkenburg}}}, \bibinfo {author} {\bibfnamefont {R.~A.}\ \bibnamefont
  {{Ibata}}}, \bibinfo {author} {\bibfnamefont {M.~L.~M.}\ \bibnamefont
  {{Collins}}}, \bibinfo {author} {\bibfnamefont {B.~P.~M.}\ \bibnamefont
  {{Laevens}}}, \bibinfo {author} {\bibfnamefont {D.}~\bibnamefont {{Mackey}}},
  \bibinfo {author} {\bibfnamefont {R.~M.}\ \bibnamefont {{Rich}}}, \bibinfo
  {author} {\bibfnamefont {D.~S.}\ \bibnamefont {{Aguado}}}, \bibinfo {author}
  {\bibfnamefont {A.}~\bibnamefont {{Arentsen}}}, \bibinfo {author}
  {\bibfnamefont {P.}~\bibnamefont {{Jablonka}}}, \bibinfo {author}
  {\bibfnamefont {J.~I.}\ \bibnamefont {{Gonzalez Hernandez}}}, \bibinfo
  {author} {\bibfnamefont {J.~F.}\ \bibnamefont {{Navarro}}}, \ and\ \bibinfo
  {author} {\bibfnamefont {R.}~\bibnamefont {{Sanchez-Janssen}}},\ }\bibfield
  {title} {\enquote {\bibinfo {title} {{The Pristine Dwarf-Galaxy survey - II.
  In-depth observational study of the faint Milky Way satellite Sagittarius
  II}},}\ }\href@noop {} {\bibfield  {journal} {\bibinfo  {journal} {arXiv
  e-prints}\ } (\bibinfo {year} {2019})},\ \Eprint
  {http://arxiv.org/abs/1902.02780} {arXiv:1902.02780} \BibitemShut {NoStop}%
\bibitem [{\citenamefont {{Niederste-Ostholt}}\ \emph
  {et~al.}(2009)\citenamefont {{Niederste-Ostholt}}, \citenamefont
  {{Belokurov}}, \citenamefont {{Evans}}, \citenamefont {{Gilmore}},
  \citenamefont {{Wyse}},\ and\ \citenamefont
  {{Norris}}}]{Niederste-Ostholt2009MNRAS.398.1771N}%
  \BibitemOpen
  \bibfield  {author} {\bibinfo {author} {\bibfnamefont {M.}~\bibnamefont
  {{Niederste-Ostholt}}}, \bibinfo {author} {\bibfnamefont {V.}~\bibnamefont
  {{Belokurov}}}, \bibinfo {author} {\bibfnamefont {N.~W.}\ \bibnamefont
  {{Evans}}}, \bibinfo {author} {\bibfnamefont {G.}~\bibnamefont {{Gilmore}}},
  \bibinfo {author} {\bibfnamefont {R.~F.~G.}\ \bibnamefont {{Wyse}}}, \ and\
  \bibinfo {author} {\bibfnamefont {J.~E.}\ \bibnamefont {{Norris}}},\
  }\bibfield  {title} {\enquote {\bibinfo {title} {{The origin of Segue 1}},}\
  }\href {\doibase 10.1111/j.1365-2966.2009.15287.x} {\bibfield  {journal}
  {\bibinfo  {journal} {Monthly Notices of the Royal Astronomical Society}\
  }\textbf {\bibinfo {volume} {398}},\ \bibinfo {pages} {1771--1781} (\bibinfo
  {year} {2009})},\ \Eprint {http://arxiv.org/abs/0906.3669} {arXiv:0906.3669
  [astro-ph.GA]} \BibitemShut {NoStop}%
\bibitem [{\citenamefont {{Simon}}\ \emph {et~al.}(2011)\citenamefont
  {{Simon}}, \citenamefont {{Geha}}, \citenamefont {{Minor}}, \citenamefont
  {{Martinez}}, \citenamefont {{Kirby}}, \citenamefont {{Bullock}},
  \citenamefont {{Kaplinghat}}, \citenamefont {{Strigari}}, \citenamefont
  {{Willman}}, \citenamefont {{Choi}}, \citenamefont {{Tollerud}},\ and\
  \citenamefont {{Wolf}}}]{Simon2011ApJ...733...46S}%
  \BibitemOpen
  \bibfield  {author} {\bibinfo {author} {\bibfnamefont {J.~D.}\ \bibnamefont
  {{Simon}}}, \bibinfo {author} {\bibfnamefont {M.}~\bibnamefont {{Geha}}},
  \bibinfo {author} {\bibfnamefont {Q.~E.}\ \bibnamefont {{Minor}}}, \bibinfo
  {author} {\bibfnamefont {G.~D.}\ \bibnamefont {{Martinez}}}, \bibinfo
  {author} {\bibfnamefont {E.~N.}\ \bibnamefont {{Kirby}}}, \bibinfo {author}
  {\bibfnamefont {J.~S.}\ \bibnamefont {{Bullock}}}, \bibinfo {author}
  {\bibfnamefont {M.}~\bibnamefont {{Kaplinghat}}}, \bibinfo {author}
  {\bibfnamefont {L.~E.}\ \bibnamefont {{Strigari}}}, \bibinfo {author}
  {\bibfnamefont {B.}~\bibnamefont {{Willman}}}, \bibinfo {author}
  {\bibfnamefont {P.~I.}\ \bibnamefont {{Choi}}}, \bibinfo {author}
  {\bibfnamefont {E.~J.}\ \bibnamefont {{Tollerud}}}, \ and\ \bibinfo {author}
  {\bibfnamefont {J.}~\bibnamefont {{Wolf}}},\ }\bibfield  {title} {\enquote
  {\bibinfo {title} {{A Complete Spectroscopic Survey of the Milky Way
  Satellite Segue 1: The Darkest Galaxy}},}\ }\href {\doibase
  10.1088/0004-637X/733/1/46} {\bibfield  {journal} {\bibinfo  {journal} {ApJ}\
  }\textbf {\bibinfo {volume} {733}},\ \bibinfo {pages} {46--+} (\bibinfo
  {year} {2011})},\ \Eprint {http://arxiv.org/abs/1007.4198} {arXiv:1007.4198
  [astro-ph.GA]} \BibitemShut {NoStop}%
\bibitem [{\citenamefont {{Siegel}}\ \emph {et~al.}(2008)\citenamefont
  {{Siegel}}, \citenamefont {{Shetrone}},\ and\ \citenamefont
  {{Irwin}}}]{Siegel2008AJ....135.2084S}%
  \BibitemOpen
  \bibfield  {author} {\bibinfo {author} {\bibfnamefont {M.~H.}\ \bibnamefont
  {{Siegel}}}, \bibinfo {author} {\bibfnamefont {M.~D.}\ \bibnamefont
  {{Shetrone}}}, \ and\ \bibinfo {author} {\bibfnamefont {M.}~\bibnamefont
  {{Irwin}}},\ }\bibfield  {title} {\enquote {\bibinfo {title} {{Trimming Down
  the Willman 1 dSph}},}\ }\href {\doibase 10.1088/0004-6256/135/6/2084}
  {\bibfield  {journal} {\bibinfo  {journal} {The Astronomical Journal}\
  }\textbf {\bibinfo {volume} {135}},\ \bibinfo {pages} {2084--2094} (\bibinfo
  {year} {2008})},\ \Eprint {http://arxiv.org/abs/0803.2489} {arXiv:0803.2489}
  \BibitemShut {NoStop}%
\bibitem [{\citenamefont {{Willman}}\ \emph {et~al.}(2011)\citenamefont
  {{Willman}}, \citenamefont {{Geha}}, \citenamefont {{Strader}}, \citenamefont
  {{Strigari}}, \citenamefont {{Simon}}, \citenamefont {{Kirby}}, \citenamefont
  {{Ho}},\ and\ \citenamefont {{Warres}}}]{Willman2011AJ....142..128W}%
  \BibitemOpen
  \bibfield  {author} {\bibinfo {author} {\bibfnamefont {B.}~\bibnamefont
  {{Willman}}}, \bibinfo {author} {\bibfnamefont {M.}~\bibnamefont {{Geha}}},
  \bibinfo {author} {\bibfnamefont {J.}~\bibnamefont {{Strader}}}, \bibinfo
  {author} {\bibfnamefont {L.~E.}\ \bibnamefont {{Strigari}}}, \bibinfo
  {author} {\bibfnamefont {J.~D.}\ \bibnamefont {{Simon}}}, \bibinfo {author}
  {\bibfnamefont {E.}~\bibnamefont {{Kirby}}}, \bibinfo {author} {\bibfnamefont
  {N.}~\bibnamefont {{Ho}}}, \ and\ \bibinfo {author} {\bibfnamefont
  {A.}~\bibnamefont {{Warres}}},\ }\bibfield  {title} {\enquote {\bibinfo
  {title} {{Willman 1 - A Probable Dwarf Galaxy with an Irregular Kinematic
  Distribution}},}\ }\href {\doibase 10.1088/0004-6256/142/4/128} {\bibfield
  {journal} {\bibinfo  {journal} {The Astronomical Journal}\ }\textbf {\bibinfo
  {volume} {142}},\ \bibinfo {eid} {128} (\bibinfo {year} {2011})},\ \Eprint
  {http://arxiv.org/abs/1007.3499} {arXiv:1007.3499} \BibitemShut {NoStop}%
\bibitem [{\citenamefont {{Torrealba}}\ \emph {et~al.}(2016)\citenamefont
  {{Torrealba}}, \citenamefont {{Koposov}}, \citenamefont {{Belokurov}},\ and\
  \citenamefont {{Irwin}}}]{Torrealba2016MNRAS.459.2370T}%
  \BibitemOpen
  \bibfield  {author} {\bibinfo {author} {\bibfnamefont {G.}~\bibnamefont
  {{Torrealba}}}, \bibinfo {author} {\bibfnamefont {S.~E.}\ \bibnamefont
  {{Koposov}}}, \bibinfo {author} {\bibfnamefont {V.}~\bibnamefont
  {{Belokurov}}}, \ and\ \bibinfo {author} {\bibfnamefont {M.}~\bibnamefont
  {{Irwin}}},\ }\bibfield  {title} {\enquote {\bibinfo {title} {{The feeble
  giant. Discovery of a large and diffuse Milky Way dwarf galaxy in the
  constellation of Crater}},}\ }\href {\doibase 10.1093/mnras/stw733}
  {\bibfield  {journal} {\bibinfo  {journal} {Monthly Notices of the Royal
  Astronomical Society}\ }\textbf {\bibinfo {volume} {459}},\ \bibinfo {pages}
  {2370--2378} (\bibinfo {year} {2016})},\ \Eprint
  {http://arxiv.org/abs/1601.07178} {arXiv:1601.07178 [astro-ph.GA]}
  \BibitemShut {NoStop}%
\bibitem [{\citenamefont {{Pace}}\ \emph {et~al.}(2020)\citenamefont {{Pace}},
  \citenamefont {{Kaplinghat}}, \citenamefont {{Kirby}}, \citenamefont
  {{Simon}}, \citenamefont {{Tollerud}}, \citenamefont {{Mu{\~n}oz}},
  \citenamefont {{C{\^o}t{\'e}}}, \citenamefont {{Djorgovski}},\ and\
  \citenamefont {{Geha}}}]{Pace2020arXiv200209503P}%
  \BibitemOpen
  \bibfield  {author} {\bibinfo {author} {\bibfnamefont {Andrew~B.}\
  \bibnamefont {{Pace}}}, \bibinfo {author} {\bibfnamefont {Manoj}\
  \bibnamefont {{Kaplinghat}}}, \bibinfo {author} {\bibfnamefont {Evan}\
  \bibnamefont {{Kirby}}}, \bibinfo {author} {\bibfnamefont {Joshua~D.}\
  \bibnamefont {{Simon}}}, \bibinfo {author} {\bibfnamefont {Erik}\
  \bibnamefont {{Tollerud}}}, \bibinfo {author} {\bibfnamefont {Ricardo~R.}\
  \bibnamefont {{Mu{\~n}oz}}}, \bibinfo {author} {\bibfnamefont {Patrick}\
  \bibnamefont {{C{\^o}t{\'e}}}}, \bibinfo {author} {\bibfnamefont {S.~G.}\
  \bibnamefont {{Djorgovski}}}, \ and\ \bibinfo {author} {\bibfnamefont
  {Marla}\ \bibnamefont {{Geha}}},\ }\bibfield  {title} {\enquote {\bibinfo
  {title} {{Multiple chemodynamic stellar populations of the Ursa Minor dwarf
  spheroidal galaxy}},}\ }\href {\doibase 10.1093/mnras/staa1419} {\bibfield
  {journal} {\bibinfo  {journal} {Monthly Notices of the Royal Astronomical
  Society}\ }\textbf {\bibinfo {volume} {495}},\ \bibinfo {pages} {3022--3040}
  (\bibinfo {year} {2020})},\ \Eprint {http://arxiv.org/abs/2002.09503}
  {arXiv:2002.09503 [astro-ph.GA]} \BibitemShut {NoStop}%
\bibitem [{\citenamefont {{Gaia Collaboration}}\ \emph
  {et~al.}(2018{\natexlab{a}})\citenamefont {{Gaia Collaboration}},
  \citenamefont {{Brown}}, \citenamefont {{Vallenari}}, \citenamefont
  {{Prusti}}, \citenamefont {{de Bruijne}}, \citenamefont {{Babusiaux}},
  \citenamefont {{Bailer-Jones}}, \citenamefont {{Biermann}}, \citenamefont
  {{Evans}}, \citenamefont {{Eyer}},\ and\ \citenamefont
  {et~al.}}]{GaiaBrown2018A&A...616A...1G}%
  \BibitemOpen
  \bibfield  {author} {\bibinfo {author} {\bibnamefont {{Gaia Collaboration}}},
  \bibinfo {author} {\bibfnamefont {A.~G.~A.}\ \bibnamefont {{Brown}}},
  \bibinfo {author} {\bibfnamefont {A.}~\bibnamefont {{Vallenari}}}, \bibinfo
  {author} {\bibfnamefont {T.}~\bibnamefont {{Prusti}}}, \bibinfo {author}
  {\bibfnamefont {J.~H.~J.}\ \bibnamefont {{de Bruijne}}}, \bibinfo {author}
  {\bibfnamefont {C.}~\bibnamefont {{Babusiaux}}}, \bibinfo {author}
  {\bibfnamefont {C.~A.~L.}\ \bibnamefont {{Bailer-Jones}}}, \bibinfo {author}
  {\bibfnamefont {M.}~\bibnamefont {{Biermann}}}, \bibinfo {author}
  {\bibfnamefont {D.~W.}\ \bibnamefont {{Evans}}}, \bibinfo {author}
  {\bibfnamefont {L.}~\bibnamefont {{Eyer}}}, \ and\ \bibinfo {author}
  {\bibnamefont {et~al.}},\ }\bibfield  {title} {\enquote {\bibinfo {title}
  {{Gaia Data Release 2. Summary of the contents and survey properties}},}\
  }\href {\doibase 10.1051/0004-6361/201833051} {\bibfield  {journal} {\bibinfo
   {journal} {AAP}\ }\textbf {\bibinfo {volume} {616}},\ \bibinfo {eid} {A1}
  (\bibinfo {year} {2018}{\natexlab{a}})},\ \Eprint
  {http://arxiv.org/abs/1804.09365} {arXiv:1804.09365} \BibitemShut {NoStop}%
\bibitem [{\citenamefont {{Gaia Collaboration}}\ \emph
  {et~al.}(2018{\natexlab{b}})\citenamefont {{Gaia Collaboration}},
  \citenamefont {{Helmi}}, \citenamefont {{van Leeuwen}}, \citenamefont
  {{McMillan}}, \citenamefont {{Massari}}, \citenamefont {{Antoja}},
  \citenamefont {{Robin}}, \citenamefont {{Lindegren}}, \citenamefont
  {{Bastian}}, \citenamefont {{Arenou}},\ and\ \citenamefont
  {et~al.}}]{GaiaHelmi2018A&A...616A..12G}%
  \BibitemOpen
  \bibfield  {author} {\bibinfo {author} {\bibnamefont {{Gaia Collaboration}}},
  \bibinfo {author} {\bibfnamefont {A.}~\bibnamefont {{Helmi}}}, \bibinfo
  {author} {\bibfnamefont {F.}~\bibnamefont {{van Leeuwen}}}, \bibinfo {author}
  {\bibfnamefont {P.~J.}\ \bibnamefont {{McMillan}}}, \bibinfo {author}
  {\bibfnamefont {D.}~\bibnamefont {{Massari}}}, \bibinfo {author}
  {\bibfnamefont {T.}~\bibnamefont {{Antoja}}}, \bibinfo {author}
  {\bibfnamefont {A.~C.}\ \bibnamefont {{Robin}}}, \bibinfo {author}
  {\bibfnamefont {L.}~\bibnamefont {{Lindegren}}}, \bibinfo {author}
  {\bibfnamefont {U.}~\bibnamefont {{Bastian}}}, \bibinfo {author}
  {\bibfnamefont {F.}~\bibnamefont {{Arenou}}}, \ and\ \bibinfo {author}
  {\bibnamefont {et~al.}},\ }\bibfield  {title} {\enquote {\bibinfo {title}
  {{Gaia Data Release 2. Kinematics of globular clusters and dwarf galaxies
  around the Milky Way}},}\ }\href {\doibase 10.1051/0004-6361/201832698}
  {\bibfield  {journal} {\bibinfo  {journal} {AAP}\ }\textbf {\bibinfo {volume}
  {616}},\ \bibinfo {eid} {A12} (\bibinfo {year} {2018}{\natexlab{b}})},\
  \Eprint {http://arxiv.org/abs/1804.09381} {arXiv:1804.09381} \BibitemShut
  {NoStop}%
\bibitem [{\citenamefont {{Geringer-Sameth}}\ \emph
  {et~al.}(2015{\natexlab{b}})\citenamefont {{Geringer-Sameth}}, \citenamefont
  {Koushiappas},\ and\ \citenamefont {Walker}}]{Geringer-Sameth:2014qqa}%
  \BibitemOpen
  \bibfield  {author} {\bibinfo {author} {\bibfnamefont {Alex}\ \bibnamefont
  {{Geringer-Sameth}}}, \bibinfo {author} {\bibfnamefont {Savvas~M.}\
  \bibnamefont {Koushiappas}}, \ and\ \bibinfo {author} {\bibfnamefont
  {Matthew~G.}\ \bibnamefont {Walker}},\ }\bibfield  {title} {\enquote
  {\bibinfo {title} {A {{Comprehensive Search}} for {{Dark Matter
  Annihilation}} in {{Dwarf Galaxies}}},}\ }\href {\doibase
  10.1103/PhysRevD.91.083535} {\bibfield  {journal} {\bibinfo  {journal} {Phys.
  Rev. D}\ }\textbf {\bibinfo {volume} {91}},\ \bibinfo {pages} {083535}
  (\bibinfo {year} {2015}{\natexlab{b}})},\ \Eprint
  {http://arxiv.org/abs/1410.2242} {arXiv:1410.2242} \BibitemShut {NoStop}%
\bibitem [{\citenamefont {{Walker}}\ \emph {et~al.}(2016)\citenamefont
  {{Walker}}, \citenamefont {{Mateo}}, \citenamefont {{Olszewski}},
  \citenamefont {{Koposov}}, \citenamefont {{Belokurov}}, \citenamefont
  {{Jethwa}}, \citenamefont {{Nidever}}, \citenamefont {{Bonnivard}},
  \citenamefont {{Bailey}}, \citenamefont {{Bell}},\ and\ \citenamefont
  {{Loebman}}}]{Walker:2016adk}%
  \BibitemOpen
  \bibfield  {author} {\bibinfo {author} {\bibfnamefont {Matthew~G.}\
  \bibnamefont {{Walker}}}, \bibinfo {author} {\bibfnamefont {Mario}\
  \bibnamefont {{Mateo}}}, \bibinfo {author} {\bibfnamefont {Edward~W.}\
  \bibnamefont {{Olszewski}}}, \bibinfo {author} {\bibfnamefont {Sergey}\
  \bibnamefont {{Koposov}}}, \bibinfo {author} {\bibfnamefont {Vasily}\
  \bibnamefont {{Belokurov}}}, \bibinfo {author} {\bibfnamefont {Prashin}\
  \bibnamefont {{Jethwa}}}, \bibinfo {author} {\bibfnamefont {David~L.}\
  \bibnamefont {{Nidever}}}, \bibinfo {author} {\bibfnamefont {Vincent}\
  \bibnamefont {{Bonnivard}}}, \bibinfo {author} {\bibfnamefont {III}\
  \bibnamefont {{Bailey}}, \bibfnamefont {John~I.}}, \bibinfo {author}
  {\bibfnamefont {Eric~F.}\ \bibnamefont {{Bell}}}, \ and\ \bibinfo {author}
  {\bibfnamefont {Sarah~R.}\ \bibnamefont {{Loebman}}},\ }\bibfield  {title}
  {\enquote {\bibinfo {title} {{Magellan/M2FS Spectroscopy of Tucana 2 and Grus
  1}},}\ }\href {\doibase 10.3847/0004-637X/819/1/53} {\bibfield  {journal}
  {\bibinfo  {journal} {ApJ}\ }\textbf {\bibinfo {volume} {819}},\ \bibinfo
  {eid} {53} (\bibinfo {year} {2016})},\ \Eprint
  {http://arxiv.org/abs/1511.06296} {arXiv:1511.06296 [astro-ph.GA]}
  \BibitemShut {NoStop}%
\bibitem [{\citenamefont {Bruel}\ \emph {et~al.}(2018)\citenamefont {Bruel},
  \citenamefont {Burnett}, \citenamefont {Digel}, \citenamefont {Johannesson},
  \citenamefont {Omodei},\ and\ \citenamefont {Wood}}]{Bruel:2018lac}%
  \BibitemOpen
  \bibfield  {author} {\bibinfo {author} {\bibfnamefont {P.}~\bibnamefont
  {Bruel}}, \bibinfo {author} {\bibfnamefont {T.~H.}\ \bibnamefont {Burnett}},
  \bibinfo {author} {\bibfnamefont {S.~W.}\ \bibnamefont {Digel}}, \bibinfo
  {author} {\bibfnamefont {G.}~\bibnamefont {Johannesson}}, \bibinfo {author}
  {\bibfnamefont {N.}~\bibnamefont {Omodei}}, \ and\ \bibinfo {author}
  {\bibfnamefont {M.}~\bibnamefont {Wood}} (\bibinfo {collaboration}
  {Fermi-LAT}),\ }\bibfield  {title} {\enquote {\bibinfo {title} {{Fermi-LAT
  improved Pass~8 event selection}},}\ }in\ \href@noop {} {\emph {\bibinfo
  {booktitle} {{8th International Fermi Symposium: Celebrating 10 Year of Fermi
  Baltimore, Maryland, USA, October 14-19, 2018}}}}\ (\bibinfo {year} {2018})\
  \Eprint {http://arxiv.org/abs/1810.11394} {arXiv:1810.11394 [astro-ph.IM]}
  \BibitemShut {NoStop}%
\bibitem [{\citenamefont {Cirelli}\ \emph {et~al.}(2011)\citenamefont
  {Cirelli}, \citenamefont {Corcella}, \citenamefont {Hektor}, \citenamefont
  {Hutsi}, \citenamefont {Kadastik}, \citenamefont {Panci}, \citenamefont
  {Raidal}, \citenamefont {Sala},\ and\ \citenamefont
  {Strumia}}]{Cirelli:2010xx}%
  \BibitemOpen
  \bibfield  {author} {\bibinfo {author} {\bibfnamefont {Marco}\ \bibnamefont
  {Cirelli}}, \bibinfo {author} {\bibfnamefont {Gennaro}\ \bibnamefont
  {Corcella}}, \bibinfo {author} {\bibfnamefont {Andi}\ \bibnamefont {Hektor}},
  \bibinfo {author} {\bibfnamefont {Gert}\ \bibnamefont {Hutsi}}, \bibinfo
  {author} {\bibfnamefont {Mario}\ \bibnamefont {Kadastik}}, \bibinfo {author}
  {\bibfnamefont {Paolo}\ \bibnamefont {Panci}}, \bibinfo {author}
  {\bibfnamefont {Martti}\ \bibnamefont {Raidal}}, \bibinfo {author}
  {\bibfnamefont {Filippo}\ \bibnamefont {Sala}}, \ and\ \bibinfo {author}
  {\bibfnamefont {Alessandro}\ \bibnamefont {Strumia}},\ }\bibfield  {title}
  {\enquote {\bibinfo {title} {{PPPC 4 DM ID: A Poor Particle Physicist
  Cookbook for Dark Matter Indirect Detection}},}\ }\href {\doibase
  10.1088/1475-7516/2012/10/E01, 10.1088/1475-7516/2011/03/051} {\bibfield
  {journal} {\bibinfo  {journal} {JCAP}\ }\textbf {\bibinfo {volume} {1103}},\
  \bibinfo {pages} {051} (\bibinfo {year} {2011})},\ \bibinfo {note} {[Erratum:
  JCAP1210,E01(2012)]},\ \Eprint {http://arxiv.org/abs/1012.4515}
  {arXiv:1012.4515 [hep-ph]} \BibitemShut {NoStop}%
\bibitem [{\citenamefont {Collaboration}(2015)}]{Ackermann:2015zua}%
  \BibitemOpen
  \bibfield  {author} {\bibinfo {author} {\bibfnamefont {Fermi-LAT}\
  \bibnamefont {Collaboration}},\ }\bibfield  {title} {\enquote {\bibinfo
  {title} {Searching for {{Dark Matter Annihilation}} from {{Milky Way Dwarf
  Spheroidal Galaxies}} with {{Six Years}} of {{Fermi}}-{{LAT Data}}},}\ }\href
  {\doibase 10.1103/PhysRevLett.115.231301} {\bibfield  {journal} {\bibinfo
  {journal} {Phys. Rev. Lett.}\ }\textbf {\bibinfo {volume} {115}},\ \bibinfo
  {pages} {231301} (\bibinfo {year} {2015})},\ \Eprint
  {http://arxiv.org/abs/1503.02641} {arXiv:1503.02641} \BibitemShut {NoStop}%
\bibitem [{\citenamefont {{Evans}}\ \emph {et~al.}(2016)\citenamefont
  {{Evans}}, \citenamefont {{Sanders}},\ and\ \citenamefont
  {{Geringer-Sameth}}}]{Evans2016PhRvD..93j3512E}%
  \BibitemOpen
  \bibfield  {author} {\bibinfo {author} {\bibfnamefont {N.~W.}\ \bibnamefont
  {{Evans}}}, \bibinfo {author} {\bibfnamefont {J.~L.}\ \bibnamefont
  {{Sanders}}}, \ and\ \bibinfo {author} {\bibfnamefont {A.}~\bibnamefont
  {{Geringer-Sameth}}},\ }\bibfield  {title} {\enquote {\bibinfo {title}
  {{Simple J-factors and D-factors for indirect dark matter detection}},}\
  }\href {\doibase 10.1103/PhysRevD.93.103512} {\bibfield  {journal} {\bibinfo
  {journal} {PRD}\ }\textbf {\bibinfo {volume} {93}},\ \bibinfo {eid} {103512}
  (\bibinfo {year} {2016})},\ \Eprint {http://arxiv.org/abs/1604.05599}
  {arXiv:1604.05599} \BibitemShut {NoStop}%
\bibitem [{\citenamefont {{Sanders}}\ \emph {et~al.}(2016)\citenamefont
  {{Sanders}}, \citenamefont {{Evans}}, \citenamefont {{Geringer-Sameth}},\
  and\ \citenamefont {{Dehnen}}}]{Sanders2016PhRvD..94f3521S}%
  \BibitemOpen
  \bibfield  {author} {\bibinfo {author} {\bibfnamefont {J.~L.}\ \bibnamefont
  {{Sanders}}}, \bibinfo {author} {\bibfnamefont {N.~W.}\ \bibnamefont
  {{Evans}}}, \bibinfo {author} {\bibfnamefont {A.}~\bibnamefont
  {{Geringer-Sameth}}}, \ and\ \bibinfo {author} {\bibfnamefont
  {W.}~\bibnamefont {{Dehnen}}},\ }\bibfield  {title} {\enquote {\bibinfo
  {title} {{Indirect dark matter detection for flattened dwarf galaxies}},}\
  }\href {\doibase 10.1103/PhysRevD.94.063521} {\bibfield  {journal} {\bibinfo
  {journal} {PRD}\ }\textbf {\bibinfo {volume} {94}},\ \bibinfo {eid} {063521}
  (\bibinfo {year} {2016})},\ \Eprint {http://arxiv.org/abs/1604.05493}
  {arXiv:1604.05493} \BibitemShut {NoStop}%
\bibitem [{\citenamefont {{Hayashi}}\ \emph {et~al.}(2016)\citenamefont
  {{Hayashi}}, \citenamefont {{Ichikawa}}, \citenamefont {{Matsumoto}},
  \citenamefont {{Ibe}}, \citenamefont {{Ishigaki}},\ and\ \citenamefont
  {{Sugai}}}]{Hayashi2016MNRAS.461.2914H}%
  \BibitemOpen
  \bibfield  {author} {\bibinfo {author} {\bibfnamefont {K.}~\bibnamefont
  {{Hayashi}}}, \bibinfo {author} {\bibfnamefont {K.}~\bibnamefont
  {{Ichikawa}}}, \bibinfo {author} {\bibfnamefont {S.}~\bibnamefont
  {{Matsumoto}}}, \bibinfo {author} {\bibfnamefont {M.}~\bibnamefont {{Ibe}}},
  \bibinfo {author} {\bibfnamefont {M.~N.}\ \bibnamefont {{Ishigaki}}}, \ and\
  \bibinfo {author} {\bibfnamefont {H.}~\bibnamefont {{Sugai}}},\ }\bibfield
  {title} {\enquote {\bibinfo {title} {{Dark matter annihilation and decay from
  non-spherical dark halos in galactic dwarf satellites}},}\ }\href {\doibase
  10.1093/mnras/stw1457} {\bibfield  {journal} {\bibinfo  {journal} {Monthly
  Notices of the Royal Astronomical Society}\ }\textbf {\bibinfo {volume}
  {461}},\ \bibinfo {pages} {2914--2928} (\bibinfo {year} {2016})},\ \Eprint
  {http://arxiv.org/abs/1603.08046} {arXiv:1603.08046} \BibitemShut {NoStop}%
\bibitem [{\citenamefont {{Klop}}\ \emph {et~al.}(2017)\citenamefont {{Klop}},
  \citenamefont {{Zandanel}}, \citenamefont {{Hayashi}},\ and\ \citenamefont
  {{Ando}}}]{Klop2017PhRvD..95l3012K}%
  \BibitemOpen
  \bibfield  {author} {\bibinfo {author} {\bibfnamefont {Niki}\ \bibnamefont
  {{Klop}}}, \bibinfo {author} {\bibfnamefont {Fabio}\ \bibnamefont
  {{Zandanel}}}, \bibinfo {author} {\bibfnamefont {Kohei}\ \bibnamefont
  {{Hayashi}}}, \ and\ \bibinfo {author} {\bibfnamefont {Shin'ichiro}\
  \bibnamefont {{Ando}}},\ }\bibfield  {title} {\enquote {\bibinfo {title}
  {{Impact of axisymmetric mass models for dwarf spheroidal galaxies on
  indirect dark matter searches}},}\ }\href {\doibase
  10.1103/PhysRevD.95.123012} {\bibfield  {journal} {\bibinfo  {journal} {PRD}\
  }\textbf {\bibinfo {volume} {95}},\ \bibinfo {eid} {123012} (\bibinfo {year}
  {2017})},\ \Eprint {http://arxiv.org/abs/1609.03509} {arXiv:1609.03509
  [astro-ph.CO]} \BibitemShut {NoStop}%
\bibitem [{\citenamefont {Finkbeiner}\ \emph {et~al.}(2011)\citenamefont
  {Finkbeiner}, \citenamefont {Goodenough}, \citenamefont {Slatyer},
  \citenamefont {Vogelsberger},\ and\ \citenamefont
  {Weiner}}]{Finkbeiner:2010sm}%
  \BibitemOpen
  \bibfield  {author} {\bibinfo {author} {\bibfnamefont {Douglas~P.}\
  \bibnamefont {Finkbeiner}}, \bibinfo {author} {\bibfnamefont {Lisa}\
  \bibnamefont {Goodenough}}, \bibinfo {author} {\bibfnamefont {Tracy~R.}\
  \bibnamefont {Slatyer}}, \bibinfo {author} {\bibfnamefont {Mark}\
  \bibnamefont {Vogelsberger}}, \ and\ \bibinfo {author} {\bibfnamefont {Neal}\
  \bibnamefont {Weiner}},\ }\bibfield  {title} {\enquote {\bibinfo {title}
  {{Consistent Scenarios for Cosmic-Ray Excesses from Sommerfeld-Enhanced Dark
  Matter Annihilation}},}\ }\href {\doibase 10.1088/1475-7516/2011/05/002}
  {\bibfield  {journal} {\bibinfo  {journal} {JCAP}\ }\textbf {\bibinfo
  {volume} {1105}},\ \bibinfo {pages} {002} (\bibinfo {year} {2011})},\ \Eprint
  {http://arxiv.org/abs/1011.3082} {arXiv:1011.3082 [hep-ph]} \BibitemShut
  {NoStop}%
\bibitem [{\citenamefont {Hisano}\ \emph {et~al.}(2011)\citenamefont {Hisano},
  \citenamefont {Kawasaki}, \citenamefont {Kohri}, \citenamefont {Moroi},
  \citenamefont {Nakayama},\ and\ \citenamefont {Sekiguchi}}]{kohri2011}%
  \BibitemOpen
  \bibfield  {author} {\bibinfo {author} {\bibfnamefont {Junji}\ \bibnamefont
  {Hisano}}, \bibinfo {author} {\bibfnamefont {Masahiro}\ \bibnamefont
  {Kawasaki}}, \bibinfo {author} {\bibfnamefont {Kazunori}\ \bibnamefont
  {Kohri}}, \bibinfo {author} {\bibfnamefont {Takeo}\ \bibnamefont {Moroi}},
  \bibinfo {author} {\bibfnamefont {Kazunori}\ \bibnamefont {Nakayama}}, \ and\
  \bibinfo {author} {\bibfnamefont {Toyokazu}\ \bibnamefont {Sekiguchi}},\
  }\bibfield  {title} {\enquote {\bibinfo {title} {Cosmological constraints on
  dark matter models with velocity-dependent annihilation cross section},}\
  }\href {\doibase 10.1103/PhysRevD.83.123511} {\bibfield  {journal} {\bibinfo
  {journal} {Phys. Rev. D}\ }\textbf {\bibinfo {volume} {83}},\ \bibinfo
  {pages} {123511} (\bibinfo {year} {2011})},\ \Eprint
  {http://arxiv.org/abs/1102.4658} {arXiv:1102.4658} \BibitemShut {NoStop}%
\bibitem [{\citenamefont {Shelton}\ \emph {et~al.}(2015)\citenamefont
  {Shelton}, \citenamefont {Shapiro},\ and\ \citenamefont
  {Fields}}]{Shelton:2015aqa}%
  \BibitemOpen
  \bibfield  {author} {\bibinfo {author} {\bibfnamefont {Jessie}\ \bibnamefont
  {Shelton}}, \bibinfo {author} {\bibfnamefont {Stuart~L.}\ \bibnamefont
  {Shapiro}}, \ and\ \bibinfo {author} {\bibfnamefont {Brian~D.}\ \bibnamefont
  {Fields}},\ }\bibfield  {title} {\enquote {\bibinfo {title} {{Black hole
  window into $p$-wave dark matter annihilation}},}\ }\href {\doibase
  10.1103/PhysRevLett.115.231302} {\bibfield  {journal} {\bibinfo  {journal}
  {Phys. Rev. Lett.}\ }\textbf {\bibinfo {volume} {115}},\ \bibinfo {pages}
  {231302} (\bibinfo {year} {2015})},\ \Eprint
  {http://arxiv.org/abs/1506.04143} {arXiv:1506.04143 [astro-ph.HE]}
  \BibitemShut {NoStop}%
\bibitem [{\citenamefont {Sandick}\ \emph {et~al.}(2018)\citenamefont
  {Sandick}, \citenamefont {Sinha},\ and\ \citenamefont
  {Yamamoto}}]{Sandick:2016zeg}%
  \BibitemOpen
  \bibfield  {author} {\bibinfo {author} {\bibfnamefont {Pearl}\ \bibnamefont
  {Sandick}}, \bibinfo {author} {\bibfnamefont {Kuver}\ \bibnamefont {Sinha}},
  \ and\ \bibinfo {author} {\bibfnamefont {Takahiro}\ \bibnamefont
  {Yamamoto}},\ }\bibfield  {title} {\enquote {\bibinfo {title} {{Black Holes,
  Dark Matter Spikes, and Constraints on Simplified Models with $t$-Channel
  Mediators}},}\ }\href {\doibase 10.1103/PhysRevD.98.035004} {\bibfield
  {journal} {\bibinfo  {journal} {Phys. Rev.}\ }\textbf {\bibinfo {volume}
  {D98}},\ \bibinfo {pages} {035004} (\bibinfo {year} {2018})},\ \Eprint
  {http://arxiv.org/abs/1701.00067} {arXiv:1701.00067 [hep-ph]} \BibitemShut
  {NoStop}%
\bibitem [{\citenamefont {Johnson}\ \emph {et~al.}(2019)\citenamefont
  {Johnson}, \citenamefont {Caputo}, \citenamefont {Karwin}, \citenamefont
  {Murgia}, \citenamefont {Ritz},\ and\ \citenamefont
  {Shelton}}]{Johnson:2019hsm}%
  \BibitemOpen
  \bibfield  {author} {\bibinfo {author} {\bibfnamefont {Christian}\
  \bibnamefont {Johnson}}, \bibinfo {author} {\bibfnamefont {Regina}\
  \bibnamefont {Caputo}}, \bibinfo {author} {\bibfnamefont {Chris}\
  \bibnamefont {Karwin}}, \bibinfo {author} {\bibfnamefont {Simona}\
  \bibnamefont {Murgia}}, \bibinfo {author} {\bibfnamefont {Steve}\
  \bibnamefont {Ritz}}, \ and\ \bibinfo {author} {\bibfnamefont {Jessie}\
  \bibnamefont {Shelton}},\ }\bibfield  {title} {\enquote {\bibinfo {title}
  {{Search for gamma-ray emission from $p$-wave dark matter annihilation in the
  Galactic Center}},}\ }\href {\doibase 10.1103/PhysRevD.99.103007} {\bibfield
  {journal} {\bibinfo  {journal} {Phys. Rev.}\ }\textbf {\bibinfo {volume}
  {D99}},\ \bibinfo {pages} {103007} (\bibinfo {year} {2019})},\ \Eprint
  {http://arxiv.org/abs/1904.06261} {arXiv:1904.06261 [astro-ph.HE]}
  \BibitemShut {NoStop}%
\bibitem [{\citenamefont {Liu}\ \emph {et~al.}(2016)\citenamefont {Liu},
  \citenamefont {Slatyer},\ and\ \citenamefont {Zavala}}]{Liu:2016cnk}%
  \BibitemOpen
  \bibfield  {author} {\bibinfo {author} {\bibfnamefont {Hongwan}\ \bibnamefont
  {Liu}}, \bibinfo {author} {\bibfnamefont {Tracy~R.}\ \bibnamefont {Slatyer}},
  \ and\ \bibinfo {author} {\bibfnamefont {Jesús}\ \bibnamefont {Zavala}},\
  }\bibfield  {title} {\enquote {\bibinfo {title} {{Contributions to cosmic
  reionization from dark matter annihilation and decay}},}\ }\href {\doibase
  10.1103/PhysRevD.94.063507} {\bibfield  {journal} {\bibinfo  {journal} {Phys.
  Rev.}\ }\textbf {\bibinfo {volume} {D94}},\ \bibinfo {pages} {063507}
  (\bibinfo {year} {2016})},\ \Eprint {http://arxiv.org/abs/1604.02457}
  {arXiv:1604.02457 [astro-ph.CO]} \BibitemShut {NoStop}%
\bibitem [{\citenamefont {Acharya}\ \emph {et~al.}(2018)\citenamefont {Acharya}
  \emph {et~al.}}]{Acharya:2017ttl}%
  \BibitemOpen
  \bibfield  {author} {\bibinfo {author} {\bibfnamefont {B.~S.}\ \bibnamefont
  {Acharya}} \emph {et~al.} (\bibinfo {collaboration} {CTA Consortium}),\
  }\href {\doibase 10.1142/10986} {\emph {\bibinfo {title} {{Science with the
  Cherenkov Telescope Array}}}}\ (\bibinfo  {publisher} {WSP},\ \bibinfo {year}
  {2018})\ \Eprint {http://arxiv.org/abs/1709.07997} {arXiv:1709.07997
  [astro-ph.IM]} \BibitemShut {NoStop}%
\bibitem [{\citenamefont {{Bechtol}}\ \emph {et~al.}(2015)\citenamefont
  {{Bechtol}} \emph {et~al.}}]{Bechtol2015ApJ...807...50B}%
  \BibitemOpen
  \bibfield  {author} {\bibinfo {author} {\bibfnamefont {K.}~\bibnamefont
  {{Bechtol}}} \emph {et~al.} (\bibinfo {collaboration} {The DES
  Collaboration}),\ }\bibfield  {title} {\enquote {\bibinfo {title} {{Eight New
  Milky Way Companions Discovered in First-year Dark Energy Survey Data}},}\
  }\href {\doibase 10.1088/0004-637X/807/1/50} {\bibfield  {journal} {\bibinfo
  {journal} {ApJ}\ }\textbf {\bibinfo {volume} {807}},\ \bibinfo {eid} {50}
  (\bibinfo {year} {2015})},\ \Eprint {http://arxiv.org/abs/1503.02584}
  {arXiv:1503.02584} \BibitemShut {NoStop}%
\bibitem [{\citenamefont {{Drlica-Wagner}}\ \emph {et~al.}(2015)\citenamefont
  {{Drlica-Wagner}} \emph {et~al.}}]{Drlica-Wagner2015ApJ...813..109D}%
  \BibitemOpen
  \bibfield  {author} {\bibinfo {author} {\bibfnamefont {A.}~\bibnamefont
  {{Drlica-Wagner}}} \emph {et~al.} (\bibinfo {collaboration} {The DES
  Collaboration}),\ }\bibfield  {title} {\enquote {\bibinfo {title} {{Eight
  Ultra-faint Galaxy Candidates Discovered in Year Two of the Dark Energy
  Survey}},}\ }\href {\doibase 10.1088/0004-637X/813/2/109} {\bibfield
  {journal} {\bibinfo  {journal} {ApJ}\ }\textbf {\bibinfo {volume} {813}},\
  \bibinfo {eid} {109} (\bibinfo {year} {2015})},\ \Eprint
  {http://arxiv.org/abs/1508.03622} {arXiv:1508.03622} \BibitemShut {NoStop}%
\bibitem [{\citenamefont {{Homma}}\ \emph {et~al.}(2018)\citenamefont
  {{Homma}}, \citenamefont {{Chiba}}, \citenamefont {{Okamoto}}, \citenamefont
  {{Komiyama}}, \citenamefont {{Tanaka}}, \citenamefont {{Tanaka}},
  \citenamefont {{Ishigaki}}, \citenamefont {{Hayashi}}, \citenamefont
  {{Arimoto}}, \citenamefont {{Garmilla}}, \citenamefont {{Lupton}},
  \citenamefont {{Strauss}}, \citenamefont {{Miyazaki}}, \citenamefont
  {{Wang}},\ and\ \citenamefont {{Murayama}}}]{Homma2018PASJ...70S..18H}%
  \BibitemOpen
  \bibfield  {author} {\bibinfo {author} {\bibfnamefont {Daisuke}\ \bibnamefont
  {{Homma}}}, \bibinfo {author} {\bibfnamefont {Masashi}\ \bibnamefont
  {{Chiba}}}, \bibinfo {author} {\bibfnamefont {Sakurako}\ \bibnamefont
  {{Okamoto}}}, \bibinfo {author} {\bibfnamefont {Yutaka}\ \bibnamefont
  {{Komiyama}}}, \bibinfo {author} {\bibfnamefont {Masayuki}\ \bibnamefont
  {{Tanaka}}}, \bibinfo {author} {\bibfnamefont {Mikito}\ \bibnamefont
  {{Tanaka}}}, \bibinfo {author} {\bibfnamefont {Miho~N.}\ \bibnamefont
  {{Ishigaki}}}, \bibinfo {author} {\bibfnamefont {Kohei}\ \bibnamefont
  {{Hayashi}}}, \bibinfo {author} {\bibfnamefont {Nobuo}\ \bibnamefont
  {{Arimoto}}}, \bibinfo {author} {\bibfnamefont {Jos{\'e}~A.}\ \bibnamefont
  {{Garmilla}}}, \bibinfo {author} {\bibfnamefont {Robert~H.}\ \bibnamefont
  {{Lupton}}}, \bibinfo {author} {\bibfnamefont {Michael~A.}\ \bibnamefont
  {{Strauss}}}, \bibinfo {author} {\bibfnamefont {Satoshi}\ \bibnamefont
  {{Miyazaki}}}, \bibinfo {author} {\bibfnamefont {Shiang-Yu}\ \bibnamefont
  {{Wang}}}, \ and\ \bibinfo {author} {\bibfnamefont {Hitoshi}\ \bibnamefont
  {{Murayama}}},\ }\bibfield  {title} {\enquote {\bibinfo {title} {{Searches
  for new Milky Way satellites from the first two years of data of the
  Subaru/Hyper Suprime-Cam survey: Discovery of Cetus III}},}\ }\href {\doibase
  10.1093/pasj/psx050} {\bibfield  {journal} {\bibinfo  {journal} {PASJ}\
  }\textbf {\bibinfo {volume} {70}},\ \bibinfo {eid} {S18} (\bibinfo {year}
  {2018})},\ \Eprint {http://arxiv.org/abs/1704.05977} {arXiv:1704.05977
  [astro-ph.GA]} \BibitemShut {NoStop}%
\bibitem [{\citenamefont {{Homma}}\ \emph {et~al.}(2019)\citenamefont
  {{Homma}}, \citenamefont {{Chiba}}, \citenamefont {{Komiyama}}, \citenamefont
  {{Tanaka}}, \citenamefont {{Okamoto}}, \citenamefont {{Tanaka}},
  \citenamefont {{Ishigaki}}, \citenamefont {{Hayashi}}, \citenamefont
  {{Arimoto}}, \citenamefont {{Carlsten}}, \citenamefont {{Lupton}},
  \citenamefont {{Strauss}}, \citenamefont {{Miyazaki}}, \citenamefont
  {{Torrealba}}, \citenamefont {{Wang}},\ and\ \citenamefont
  {{Murayama}}}]{Homma2019PASJ..tmp...91H}%
  \BibitemOpen
  \bibfield  {author} {\bibinfo {author} {\bibfnamefont {Daisuke}\ \bibnamefont
  {{Homma}}}, \bibinfo {author} {\bibfnamefont {Masashi}\ \bibnamefont
  {{Chiba}}}, \bibinfo {author} {\bibfnamefont {Yutaka}\ \bibnamefont
  {{Komiyama}}}, \bibinfo {author} {\bibfnamefont {Masayuki}\ \bibnamefont
  {{Tanaka}}}, \bibinfo {author} {\bibfnamefont {Sakurako}\ \bibnamefont
  {{Okamoto}}}, \bibinfo {author} {\bibfnamefont {Mikito}\ \bibnamefont
  {{Tanaka}}}, \bibinfo {author} {\bibfnamefont {Miho~N.}\ \bibnamefont
  {{Ishigaki}}}, \bibinfo {author} {\bibfnamefont {Kohei}\ \bibnamefont
  {{Hayashi}}}, \bibinfo {author} {\bibfnamefont {Nobuo}\ \bibnamefont
  {{Arimoto}}}, \bibinfo {author} {\bibfnamefont {Scott~G.}\ \bibnamefont
  {{Carlsten}}}, \bibinfo {author} {\bibfnamefont {Robert~H.}\ \bibnamefont
  {{Lupton}}}, \bibinfo {author} {\bibfnamefont {Michael~A.}\ \bibnamefont
  {{Strauss}}}, \bibinfo {author} {\bibfnamefont {Satoshi}\ \bibnamefont
  {{Miyazaki}}}, \bibinfo {author} {\bibfnamefont {Gabriel}\ \bibnamefont
  {{Torrealba}}}, \bibinfo {author} {\bibfnamefont {Shiang-Yu}\ \bibnamefont
  {{Wang}}}, \ and\ \bibinfo {author} {\bibfnamefont {Hitoshi}\ \bibnamefont
  {{Murayama}}},\ }\bibfield  {title} {\enquote {\bibinfo {title} {{Bo{\"o}tes.
  IV. A new Milky Way satellite discovered in the Subaru Hyper Suprime-Cam
  Survey and implications for the missing satellite problem}},}\ }\href
  {\doibase 10.1093/pasj/psz076} {\bibfield  {journal} {\bibinfo  {journal}
  {PASJ}\ ,\ \bibinfo {pages} {91}} (\bibinfo {year} {2019})},\ \Eprint
  {http://arxiv.org/abs/1906.07332} {arXiv:1906.07332 [astro-ph.GA]}
  \BibitemShut {NoStop}%
\bibitem [{\citenamefont {Ivezi\'c}\ \emph {et~al.}(2008)\citenamefont
  {Ivezi\'c}, \citenamefont {Tyson}, \citenamefont {Acosta}, \citenamefont
  {Allsman}, \citenamefont {Anderson}, \citenamefont {Andrew}, \citenamefont
  {Angel}, \citenamefont {Axelrod}, \citenamefont {Barr}, \citenamefont
  {Becker} \emph {et~al.}}]{ivezic2008lsst}%
  \BibitemOpen
  \bibfield  {author} {\bibinfo {author} {\bibfnamefont {\v{Z}.}\ \bibnamefont
  {Ivezi\'c}}, \bibinfo {author} {\bibfnamefont {J.~A.}\ \bibnamefont {Tyson}},
  \bibinfo {author} {\bibfnamefont {E.}~\bibnamefont {Acosta}}, \bibinfo
  {author} {\bibfnamefont {R.}~\bibnamefont {Allsman}}, \bibinfo {author}
  {\bibfnamefont {S.~F.}\ \bibnamefont {Anderson}}, \bibinfo {author}
  {\bibfnamefont {J.}~\bibnamefont {Andrew}}, \bibinfo {author} {\bibfnamefont
  {J.~R.~P.}\ \bibnamefont {Angel}}, \bibinfo {author} {\bibfnamefont {T.~S.}\
  \bibnamefont {Axelrod}}, \bibinfo {author} {\bibfnamefont {J.~D.}\
  \bibnamefont {Barr}}, \bibinfo {author} {\bibfnamefont {A.~C.}\ \bibnamefont
  {Becker}},  \emph {et~al.},\ }\bibfield  {title} {\enquote {\bibinfo {title}
  {Lsst: from science drivers to reference design and anticipated data
  products},}\ }\href@noop {} {\bibfield  {journal} {\bibinfo  {journal} {ApJ}\
  } (\bibinfo {year} {2008})},\ \Eprint {http://arxiv.org/abs/0805.2366v4}
  {arXiv:0805.2366v4} \BibitemShut {NoStop}%
\bibitem [{\citenamefont {{Hargis}}\ \emph {et~al.}(2014)\citenamefont
  {{Hargis}}, \citenamefont {{Willman}},\ and\ \citenamefont
  {{Peter}}}]{Hargis2014ApJ...795L..13H}%
  \BibitemOpen
  \bibfield  {author} {\bibinfo {author} {\bibfnamefont {J.~R.}\ \bibnamefont
  {{Hargis}}}, \bibinfo {author} {\bibfnamefont {B.}~\bibnamefont {{Willman}}},
  \ and\ \bibinfo {author} {\bibfnamefont {A.~H.~G.}\ \bibnamefont {{Peter}}},\
  }\bibfield  {title} {\enquote {\bibinfo {title} {{Too Many, Too Few, or Just
  Right? The Predicted Number and Distribution of Milky Way Dwarf Galaxies}},}\
  }\href {\doibase 10.1088/2041-8205/795/1/L13} {\bibfield  {journal} {\bibinfo
   {journal} {ApJL}\ }\textbf {\bibinfo {volume} {795}},\ \bibinfo {eid} {L13}
  (\bibinfo {year} {2014})},\ \Eprint {http://arxiv.org/abs/1407.4470}
  {arXiv:1407.4470} \BibitemShut {NoStop}%
\bibitem [{\citenamefont {{Newton}}\ \emph {et~al.}(2018)\citenamefont
  {{Newton}}, \citenamefont {{Cautun}}, \citenamefont {{Jenkins}},
  \citenamefont {{Frenk}},\ and\ \citenamefont
  {{Helly}}}]{Newton2018MNRAS.479.2853N}%
  \BibitemOpen
  \bibfield  {author} {\bibinfo {author} {\bibfnamefont {Oliver}\ \bibnamefont
  {{Newton}}}, \bibinfo {author} {\bibfnamefont {Marius}\ \bibnamefont
  {{Cautun}}}, \bibinfo {author} {\bibfnamefont {Adrian}\ \bibnamefont
  {{Jenkins}}}, \bibinfo {author} {\bibfnamefont {Carlos~S.}\ \bibnamefont
  {{Frenk}}}, \ and\ \bibinfo {author} {\bibfnamefont {John~C.}\ \bibnamefont
  {{Helly}}},\ }\bibfield  {title} {\enquote {\bibinfo {title} {{The total
  satellite population of the Milky Way}},}\ }\href {\doibase
  10.1093/mnras/sty1085} {\bibfield  {journal} {\bibinfo  {journal} {MNRAS}\
  }\textbf {\bibinfo {volume} {479}},\ \bibinfo {pages} {2853--2870} (\bibinfo
  {year} {2018})},\ \Eprint {http://arxiv.org/abs/1708.04247} {arXiv:1708.04247
  [astro-ph.GA]} \BibitemShut {NoStop}%
\end{thebibliography}%
